\documentclass[nofootinbib,showpacs,preprintnumbers,amsmath,amssymb,floatfix]
{revtex4-2}
\usepackage{graphicx}
\usepackage{epsf}
\usepackage{wrapfig}

\begin{document}


\title{70 years of Double-Logarithmic Approximation} 

\vspace*{0.3 cm}

\author{B.I.~Ermolaev}
\affiliation{Ioffe Institute, 194021
 St.Petersburg, Russia}

\begin{abstract}
Existence of Double-Logarithmic (DL) contributions to scattering amplitudes in QED was discovered by V.V.~Sudakov in 1956 and total 
summation of DL contributions to electron-photon scattering resulted in appearance of famous Sudakov exponentials. Then, 
thanks to contributions of V.G.~Gorshkov, V.N.~Gribov, G.V.~Frolov and L.N.~Lipatov, the pattern of calculations in Double-Logarithmic  
Approximation (DLA) was constructed. Since then, DLA has become one of basic ways of describing various high-energy processes in the framework of QED, QCD and 
theory of EW  interactions. In the present paper, we remind the history of DLA and present a brief overview of application of DLA to 
various objects like form factors, scattering amplitudes, DIS structure functions. 
\end{abstract}

\pacs{12.38.Cy}

\maketitle

\section{Introduction}

History of Double-Logarithmic approximation (DLA) began in 1956 when V.V.~Sudakov studied, see Ref.~\cite{sud}, 
elastic scattering of electron off 
highly virtual photon. He found out that the leading contributions to the elastic form factor of the electron in $n^{th}$ 
order in the coupling are 
$\sim \left(\alpha \ln^2\left(|q^2|\right)\right)^n$, with $\alpha$ and $q$ being the coupling and the photon momentum. 
Such contributions were called 
double-logarithmic (DL). They dominated other contributions at large $|q^2|$ and because of that the sum of DL terms 
 defined  
the high-energy behavior of the form factor. The approach where DL terms are collected to all orders in the coupling 
whereas other contributions are neglected is called double-logarithmic approximation (DLA). Ref.~\cite{sud} 
demonstrated that the 
DL contributions came from the infrared (IR) region of the virtual momenta phase space where 
the virtual photons were almost on-shell, i.e. virtualities $k^2_i$ of their momenta $k_i~(i = 1,2,..)$ 
were small: $k^2_i \approx 0$. In DLA, 
electron form factors had the form of falling exponentials $\sim e^{-\alpha \ln^2 |q^2|}$ both for off-shell and on-shell electrons. Such 
expressions have been addressed as Sudakov exponentials and the fact that all of them decrease when $|q^2|$ grows is 
called Sudakov suppression.  

Ref.~\cite{sud} stimulated 
interest to studying DL contributions to $2 \to 2$ high-energy scattering, including backward Compton scattering. 
However, a simple/naive implementation  
 of the recipe of Ref.~\cite{sud} for selection of the regions yielding DL contributions led to incorrect results.  
In particular,  Refs.~\cite{abr,mil} claimed that the sum of DL contributions took the form of rising exponentials in contrast to the 
falling Sudakov ones. Such fast growth contradicted the Regge theory, where any $2 \to 2$-scattering amplitude cannot rise faster 
than $s^{\Delta}$, with $s$ being the total invariant energy and $\Delta$ being intercept, $\Delta \leq 1$. 
Analysis of this contradiction was done 
in Refs.~\cite{ggfl1, ggfl2,ggfl1,ggfl2,ggfl3, ggfl4} by V.G.~Gorshkov, V.N.~Gribov, G.V. Frolov, L.N. Lipatov who considered forward and 
backward annihilation $e^+e^- \to \mu^+\mu^-$ and showed that the in addition to the Sudakov DL contributions (coming from nearly on-shell, i.e virtual photons) 
there also were DL contributions from soft virtual fermions (the latter had been overlooked in Refs.~\cite{abr,mil}).  
This was a very important result 
completing the list of potential sources of DL contributions. With both  soft photon and soft fermion DL contributions accounted for, 
asymptotics of amplitudes of forward and backward annihilation acquired   the Regge form. 

All-order summation of DL 
contributions in Ref.~\cite{sud} was based on selecting involved Feynman graphs and simplifying their integrands in essential integration 
regions while Refs.~\cite{ggfl1,ggfl2} suggested quite complicated methods for constructing equations the amplitudes obey, 
so total resummation of DLs was an uneasy task. However even in low orders,  
DL contributions proved to be important. For instance,  they played crucial role
in production of narrow resonances in colliding $e^+ e^-$ beams as was 
shown in Ref.~\cite{khoze}.  

Constructing convenient methods of all-order summations of DL contribution exploited the important observation by V.N.~Gribov known as the Gribov bremsstrahlung theorem\cite{g} 
where it was shown that emission of bremsstrahlung photons in radiative processes at high energies can be factorized 
out of the scattering amplitude providing that the transverse momentum $k_{\perp}$ 
of the photon was small compared to the mass/energy scale of the process. Extension of this result to 
factorization of soft gluons  
and supplementing it with a similar condition for factorization of quarks was made in Ref.~\cite{kl}  by R.~Kirschner and L.N.~Lipatov.   
This result made it possible to convert the rather complicated approach of Refs.~\cite{ggfl1,ggfl2} in a much simpler 
and productive method to calculate elastic $2 \to 2$-processes in QCD with DL accuracy. The pattern 
suggested in Ref.~\cite{kl} is the core of the method of constructing equations 
tracing evolution with respect to the IR cut-off: Infra-Red Evolution Equations (IREE) for calculating various objects in QED, 
QCD and Electroweak interactions\footnote{The name IREE of the method was suggested by M.~Krawczyk.}. 

Note that the form factors 
studied in Ref.~\cite{sud} as well as the amplitudes in Refs.~\cite{ ggfl1, ggfl2,ggfl1,ggfl2,ggfl3, ggfl4,g,kl} depend on one variable: the invariant total energy $s$, so DLA for them sums contributions $\sim (\alpha_s \ln^2 s)^n$. In contrast, there are objects in high-energy physics which depend on several variables and where the IREE-technique suggested in Ref.~\cite{kl} requires essential development: for example, the DIS structure functions depend on  both $x$ and the photon virtuality $Q^2$, so DLA for them must include logarithms of both $x$ and 
$Q^2$. Of course, DGLAP\cite{dglap} deals with both the $x$ and $Q^2$ -dependence but does not account for 
summation of DL contributions which are quite 
important at small $x$. This gap was filled in Refs.~\cite{emr,berns,bers}.  Then, 
amplitudes of gluon production in backward quark-antiquark annihilation are expressed through multi-Reggeon 
ensemble (see e.g. \cite{el}).  Besides, the IREE method was enriched by accounting for the running coupling effects 
(see e.g. \cite{egtalpha}). 

All in all, DLA by now has become one of basic means for theoretical investigation of various reactions in 
QED, QCD and Standard Model at high energies. We begin the present paper with considering DLA for elastic and inelastic  
 form factors in different theories (QED, QCD and EW interactions) and derive, presenting all details of calculations,   
relations between the cases of massless/massive and on-shell/off-shell fermions.  After that we discuss DLA for 
high-energy processes in Regge kinematics.

Our paper is organized as follows: In Sect.~II we introduce notations and remind results of Ref.~\cite{sud} for both 
on-shell and off-shell electrons. Discussing the off-shell electrons, we introduce the cases of moderate and 
deep virtualities of the electrons, which proved to be quite useful for many high-energy processes.  Sect.~III 
deals with the Sudakov form factor, where the quarks, being either massive or massless, are on-shell.  
 The case of off-shell quarks is considered in Sect.~IV. We derive in Sects.~III,IV exponentiation of the first-loop DL contributions to the Sudakov form factor by  composing IREEs. Because of that we recap 
the IREE method in Sect.~III.  
Sudakov form factor in the context of electroweak reactions is considered in Sect.~V. We start with re-deriving the result of 
Ref.~\cite{flmm} and then account for the fermion mass corrections. Inelastic Sudakov form factor is considered in Sect.~VI 
while the second electron/quark form factor in DLA is considered in Sect.~VII.  Application of DLA to processes in 
Regge kinematics is considered in Sect.~VIII. Finally, Sect.~IX is for concluding remarks.

\section{Sudakov form factor in QED}

In this Sect. we first reproduce the reasoning of Ref.~\cite{sud} and then consider Sudakov form factors in QCD.
%
%
Consider the amplitude $A $ of scattering of an external photon with momentum $q$ off an electron: 

\begin{equation}
\label{agen} 
A (p_1,p_2,q) = e\; l_{\mu}(q) \Gamma_{\mu} (q) ,  
\end{equation}
where $l_{\mu}(q)$ is the photon polarization vector and $e$ is the electron charge. Then, $p_1$ and $p_2$ denote the initial and final momenta of the electron, so that  $q= p_2 - p_1$. The electron-photon vertex $\Gamma_{\mu}$ is parameterized 
by two form factors: 

\begin{equation}\label{gammadef}
\Gamma_{\mu} (q) = \bar{u}(p_2)\left[\gamma_{\mu} \;f (q^2) - \frac{\sigma_{\mu \nu}q_{\nu}}{2 m}\;g(q^2)\right]u(p_1), 
\end{equation}
where we have used standard notations: $f(q^2)$ and $g(q^2)$ are the electron form factors, $m$ is the electron mass and 
$\sigma_{\mu \nu} = (\gamma_{\mu}\gamma_{\nu} - \gamma_{\nu}\gamma_{\mu})/2$. In the Born approximation, $f(q^2) = 1$ and $g(q^2) = 0$. 
The first-loop contributions to both the form factors come from the first-loop contribution $\Gamma_{\mu}^{(1)}$ to $\Gamma_{\mu}$ 
as depicted in Fig.~1: 

\begin{equation}\label{gamma1}
\Gamma_{\mu}^{(1)} =  -\imath \frac{\alpha_{EM}}{4 \pi^3} \int d^4 k \frac{\bar{u}(p_2)\gamma_{\lambda} 
\left(\hat{p}_2 - \hat{k} +m\right)\gamma_{\mu} \left(\hat{p}_1 - \hat{k} +m\right) \gamma_{\lambda}u(p_1)}
{[(p_1 - k)^2 - m^2+ \imath \epsilon][(p_2 - k)^2 - m^2+ \imath \epsilon] [k^2 + \imath \epsilon]}, 
\end{equation}
where $k$ is the virtual photon momentum. 
\begin{figure}\label{sud70fig1}
\includegraphics[width=.6\textwidth]{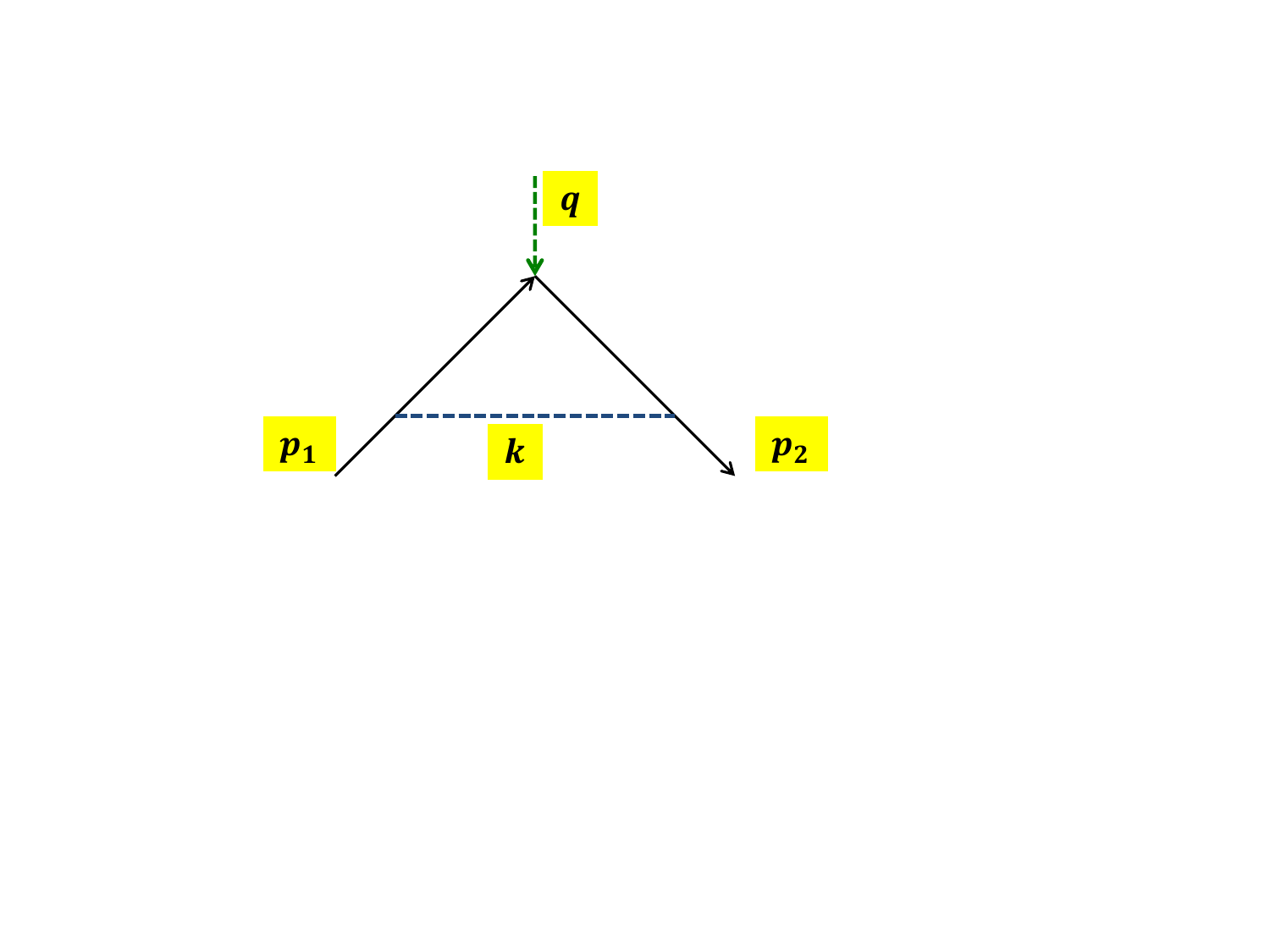}
\caption{\label{sud70fig1}
First-loop contribution to the vertex $\Gamma_{\mu}$ }
\end{figure}
We denote the electromagnetic coupling $\alpha_{EM} = 1/137$ in Eq.~(\ref{gamma1})
to avoid confusion with the Sudakov variable $\alpha$ of Eq.~(\ref{sudpar}). The integral in Eq.~(\ref{gamma1}) is one of 
the most important objects in early QED. It can be found in any textbook on QED. First, it contains 
a logarithmical divergency which is conventionally regulated with the electron charge renormailization. The finite part of the integral 
is also well-known for contributing to the anomalous magnetic momentum of electron at $q^2 = 0$. In contrast, V.V.~Sudakov calculated in Ref.~\cite{sud} the leading contribution to $\Gamma_{\mu}^{(1)}$ 
in the kinematic region, where both the photon and  electron are deeply virtual with negative $p^2_{1,2}$: 

\begin{equation}\label{sudkin}
|q^2|\gg - p^2_{1,2} \gg m^2. 
\end{equation}

In this region, the terms $\sim m, k$ can be dropped in the numerator of the integrand in Eq.~(\ref{gamma1}) as 
well as the terms $\sim m^2, k^2$ 
in the denominator. For the shortness sake we will denote $s = 2p_1p_2$ in what follows. Note that in kinematics~(\ref{sudkin}) $|q^2| = 2p_1p_2 + |p^2_2| + |p^2_1| \approx 2p_1p_2 = s$.  Integration in Eq.~(\ref{gamma1}) is especially simple when the Sudakov parametrization of $k$ is used. 
In the simplest version of the Sudakov parametrization virtualities $p^2_{1,2}$ are neglected
\footnote{A more detailed form of the parametrization, with virtualities $p^2_{1,2}$ accounted for, is considered in detail in Sect.~IV} compared to $s$: $p^2_1 \approx p^2_2 \approx 0$. Then the parametrization looks as follows: 

\begin{equation}\label{sudpar}
k = \alpha\;p_2 + \beta\;p_1 + k_{\perp},
\end{equation}
where $k_{\perp}$ denotes the component of $k$ transverse to the plane 
formed by momenta $p_{1,2}$: $k_{\perp}p_1 = k_{\perp}p_2 = 0$. The containing $k$ invariants in Eq.~(\ref{gamma1}) are 
expressed through $\alpha, \beta, k_{\perp} $: 

\begin{equation}\label{invsud}
2p_1 k = s \alpha,~2p_2k = s \beta, ~k^2 = s \alpha \beta - k^2_{\perp}
\end{equation}
and therefore

\begin{equation}\label{gamma1sud}
\Gamma_{\mu}^{(1)} \approx  -\imath  j_{\mu}\;\frac{\alpha_{EM}}{4 \pi^2} \int d \alpha d \beta d k^2_{\perp} \frac{1 }
{[s\alpha  + p^2_1 -(s\alpha \beta - k^2_{\perp}) +  \imath \epsilon][ s\beta + p^2_2- (s\alpha \beta - k^2_{\perp})  + \imath \epsilon] [s\alpha \beta - k^2_{\perp} + \imath \epsilon]}, 
\end{equation}
where we have denoted $j_{\mu} = \bar{u}(p_2)\gamma_{\mu}u(p_1)$ 
. The first and 
second factors in the denominator are IR-stable, whereas the third factor contains IR singularity, so integrating it over 
$k^2_{\perp}$ leads to dealing with the integral of the Cauchy type.  It 
 is convenient to handle such integrals,  using the  Sokhotski identity

\begin{equation}\label{soh}
\frac{1}{s \alpha \beta -k^2_{\perp} + \imath \epsilon} =  \frac{P}{s \alpha \beta -k^2_{\perp}} 
- \imath \pi \delta \left(s \alpha \beta -k^2_{\perp}\right),
\end{equation}
where the main contribution comes from the $\delta$-function. 
 The remaining integrations over $\alpha$ 
 and $\beta$ are done with logarithmic accuracy:

\begin{equation}\label{gamma1vab}
\Gamma_{\mu}^{(1)} 
\approx -  j_{\mu}\;\frac{\alpha_{EM}}{2 \pi} \int_{p^2_1/s}^1 \frac{d \alpha}{\alpha } 
\int_{p^2_2/s}^1 \frac{d \beta}{ \beta} 
= e j_{\mu} 
\left[-\frac{\alpha_{EM}}{2 \pi} \ln \left(\frac{s}{|p^2_1|}\right) \ln \left(\frac{s}{|p^2_2|}\right)\right]. 
\end{equation}

The factor in the brackets in Eq.~(\ref{gamma1vab}) is the double-logarithmic (DL) contribution to the form factor $f(s)$,   
so the sum of the Born and the first-loop contributions to $f(s)$ is 

\begin{equation}\label{f1}
 f^{(Born)} + f^{(1)}(s) = 1 - W^{(1)}(s),
\end{equation}
with the first-loop contribution  $W^{(1)}(s)$ being 

\begin{equation}\label{w1dvdef}
W^{(1)}(s) = \frac{\alpha_{EM}}{2 \pi} \ln \left(\frac{s}{|p^2_1|}\right) \ln \left(\frac{s}{|p^2_2|}\right). 
\end{equation}

Such handling the first-loop Feynman graph prompts  the following recipe for accounting for dealing with  higher-loop graphs: 
\textbf{(i)} Neglect virtual gluon momenta $k_i$ in the numerators in each expression corresponding to the Feynman graph involved and 
simplify the numerators.\\
\textbf{(ii)} Use the Sudakov parametrization for momenta of all virtual gluons and apply the identity (\ref{soh}) 
for integrating over $k_{i \perp}$. \\
\textbf{(iii)} Add up the remaining integrals over $\alpha_i, \beta_i$ corresponding to different graphs 
and make sure that they are factorized. As a result,  
the DL contribution of $n$th loop $W^{(n)}(s)$ is expressed through $W^{(1)}(s)$: 

\begin{equation}\label{wndef}
W^{(n)}(s) = \frac{(-1)^n}{n!}\; \left(W^{(1)}(s)\right)^n.
\end{equation}
and thereby we arrive at the famous Sudakov exponentiation: 

\begin{equation}\label{fdl}
f(s) = e^{-W^{(1)}(s) }. 
\end{equation}

Eq.~(\ref{fdl}) represents $f(s)$ in the double-logarithmic approximation (DLA) providing the electron is off-shell. 
The expression for $W^{(1)}(s)$ in Eq.~(\ref{w1dvdef}) does not involve any IR cut-off, so it is IR stable. However,   
Eq.~(\ref{w1dvdef}) is true for very big virtualities $p^2_{1,2}$ only. We will discuss this issue in detail in Sect.~IIIB. 
For the moment, just note that the Sudakov exponentiation with $W^{(1)}$ of Eq.~(\ref{w1dvdef}) is true when  $p^2_{1,2}$ 
obey the inequality

\begin{equation}\label{dvdef}
p^2_1 p^2_2 \gg s \mu^2,
\end{equation}
where $\mu$ is the IR cut-off and $\mu^2 \ll p^2_{1,2}$. We call such kinematic region deeply virtual (DV). In 
contrast, there is a moderately virtual (MV) kinematic region, where 

\begin{equation}\label{dvdef}
p^2_1 p^2_2 \ll s \mu^2
\end{equation}
and the first-loop contribution $W^{(1)}_{MV}(s)$ there is IR-sensitive: 

\begin{equation}\label{w1mv}
W^{(1)}_{MV}(s) = \frac{\alpha_{EM}}{4 \pi} \left[\ln^2 \left(\frac{s}{\mu^2}\right) - 
 \ln^2 \left(\frac{|p^2_1|}{\mu^2}\right)- \ln^2 \left(\frac{|p^2_2|}{\mu^2}\right)\right]. 
\end{equation}

If the electron is on-shell, i.e. $p^2_{1,2} = m^2$ and $\mu = m$, Eq.~(\ref{w1mv}) converts into the on-shell quark first-loop contribution 
$W^{(1)}_{on}(s)$: 

\begin{equation}\label{w1on}
W^{(1)}_{on}(s) = \frac{\alpha_{EM}}{4 \pi} \ln^2 \left(\frac{s}{m^2}\right). 
\end{equation}

Obviously, Eq.~(\ref{w1on}) cannot be derived from Eq.~(\ref{w1dvdef}) by putting $p^2_1 = p^2_2 =m^2$. We demonstrate  in Sect.~IIIB how to relate 
and how to exponentiate the first-loop contributions of Eqs.~(\ref{w1dvdef},\ref{w1mv},\ref{w1on}).

\section{Sudakov form factor in QCD. On-shell quarks}

Consider amplitude of scattering quark off a virtual photon with large virtuality. Similarly to QED, it involves exchange of virtual gluons. 
However, the graph-by-graph analysis presented in Ref.~\cite{sud} 
becomes really laborious for calculating 
the Sudakov form factor $F$ in QCD 
because, in contrast to QED, contributions of three-gluon vertices should be accounted for and, in addition, each vertex 
contains $SU(3)$-generators.  The form factor $f_{QCD} (s)$ was calculated in Refs.~\cite{quinn,smilga,tikt} 
with other means. Nevertheless, the result of Refs.~\cite{quinn,smilga,tikt} was surprisingly simple: 
the expression for $F (s)$ in DLA can be obtained from the QED expressions for $f (s)$ with replacement  
of the QED coupling $\alpha_{EM}$ with $\alpha_s C_F$, where   
 $C_F = (N^2-1)/2 N = 4/3$. Note  that calculations in Refs.~\cite{quinn,smilga} were based on tracing cancellation 
of IR singularities between virtual and emitted gluons whereas the starting point in Ref.~\cite{tikt} was an evolution equation of the Bethe-Salpeter type presented without derivation.  In what follows we present one more way to calculate $f_{QCD} (s)$. 
Namely, we will use the IREE approach. Our aim here is to demonstrate simplicity and efficiency of this method. We  
consider both on-shell and off-shell quarks. 

In the first place, we introduce IR cut-off $\mu$ in the transverse momentum 
  space: $k_{i \perp} \gg \mu,~(i = 1,2,...)$. Let us notice that introducing IR cut-offs inevitably causes  
  violation of the gauge invariance. Introducing $\mu$ in the transverse momentum space is equivalent to 
  compactification of the impact parameter space, so at least it does not violate the gauge invariance in the 
   longitudinal space. IREE for $F$ is depicted in Fig.~2. 
\begin{figure}\label{sud70fig2}
\includegraphics[width=.6\textwidth]{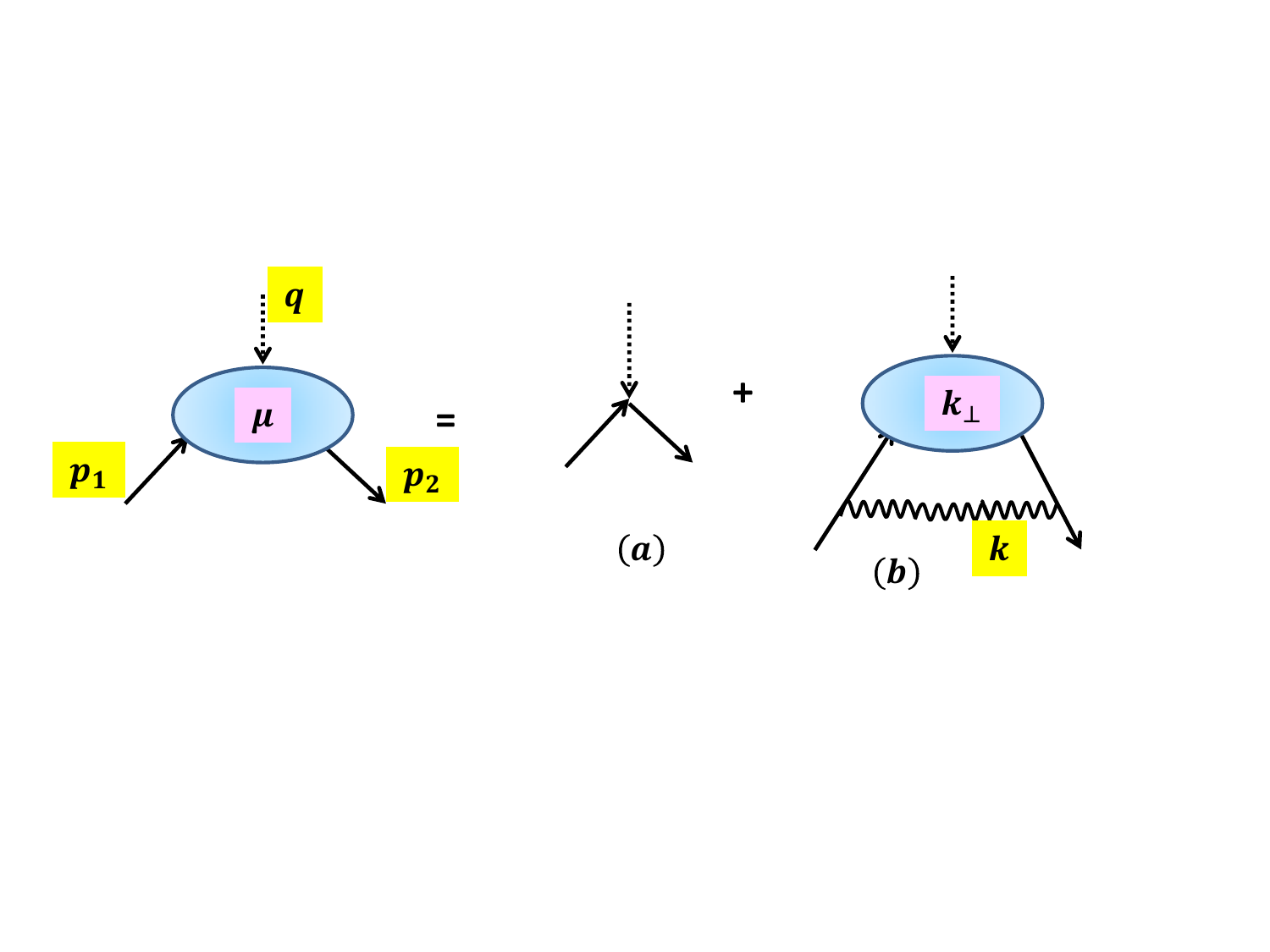}
\caption{\label{sud70fig2}
IREE for the form factor $F$ }
\end{figure}
   The blobs in Fig.~2 mean accounting for radiative corrections with DL accuracy. The letters on the 
  blobs indicate IR cut-offs. The l.h.s. of IREE in Fig.~2 is $F$ while 
  the r.h.s. consists of two terms. Graph (a) depicts the Born contribution $F^{Born}  = 1$ and 
  the graph (b) corresponds to factorization of the gluon with momentum $k$.  This gluon has minimal transverse momenta $k_{\perp}$ 
  compared to other gluons: $k_{\perp} = \min k_{i \perp}$, $(i =1,2,..)$, so according to the Gribov theorem it can be factorized, i.e. 
  its propagator can be attached to the external lines whereas $k_{\perp}$ acts as a new IR cut-off for other virtual gluons. 
  The blob on graph (b) does not depend on longitudinal Sudakov variables $\alpha$ and $\beta$, so integrations over them 
  should be done as in the first loop. In the most general case, $F$ depends on $s,p^2_1,p^2_2,m^2 $ and therefore the general 
  form of the IREE for $F$ is 
  

\begin{eqnarray}\label{eqfqcdgen}
F\left(\frac{s}{\mu^2},\frac{p^2_1}{\mu^2},\frac{p^2_2}{\mu^2},\frac{m^2}{\mu^2}\right)  
&=& 1 - \imath \frac{\alpha_s C_F}{4 \pi^2} \int  \frac{d \alpha d \beta d k^2_{\perp} }{R_1 (s \alpha, s \beta,k^2_{\perp} ) 
R_2 (s \alpha, s \beta,k^2_{\perp} ) R_3 (s \alpha, s \beta,k^2_{\perp} )},
\end{eqnarray}
where 

\begin{eqnarray}\label{r123gen}
R_1 &=& s-\alpha(1 - \beta)  + p^2_1 - k^2_{\perp} - m^2 +  \imath \epsilon,
\\ \nonumber
R_2 &=& s\beta(1 + \alpha) + p^2_2 - k^2_{\perp} -m^2 + \imath \epsilon,
\\ \nonumber
R_3 &=& s\alpha \beta - k^2_{\perp} + \imath \epsilon.
\end{eqnarray}

Integration region in Eq.~(\ref{eqfqcdgen}) depends on virtualities of the quarks and will be specified later. First, integrate Eq.~(\ref{eqfqcdgen}) over $\alpha$. 
This integration is performed with applying the Cauchy theorem. Eq.~(\ref{r123gen}) reads that the first factor, $R_1$ in the denominator of 
Eq.~(\ref{eqfqcdgen}) has a pole in $\alpha$ in the upper semi-plane at $0 < \beta <1$. The other two poles are in the lower semi-plane. 
Closing the integration contour up and taking residue in this pole, we are left with integrations over $\beta$ and $k^2_{\perp}$:

  \begin{equation}\label{eqfgen}
F\left(\frac{s}{\mu^2},\frac{p^2_1}{\mu^2},\frac{p^2_2}{\mu^2},\frac{m^2}{\mu^2}\right) = 1 - \frac{\alpha_s C_F}{2 \pi} \int_{D}  \frac{d k^2_{\perp} }{k^2_{\perp}}\;\frac{d\beta}{\beta}\; 
F\left(\frac{s}{k^2_{\perp}},\frac{p^2_1}{k^2_{\perp}},\frac{p^2_2}{k^2_{\perp}},\frac{m^2}{k^2_{\perp}}\right),
  \end{equation}
  where the integration region $D$ is, in the first place, formed by the requirement $0 < \beta <1$ 
  and also depends on the specific cases. Below we consider the following specific cases: \\
  \textbf{(A)} Massless quarks\\
  \textbf{(B) }Accounting for the quark masses.
 
 \subsection{Massless quarks}

  We start with consideration of the simplest case when the quark masses can be neglected, neglecting the quark virtualities and 
  masses in Eq.~(\ref{r123gen}) and therefore arriving at the IREE for the massless quark form factor $F_0$:

  \begin{eqnarray}\label{eqf0gen}
F_0\left(\frac{s}{\mu^2}\right)  
&=& 1 
-\imath \frac{\alpha_s C_F}{4 \pi^2} \int  \frac{d \alpha d \beta d k^2_{\perp} }
{[-s\alpha(1 - \beta) - k^2_{\perp}  +  \imath \epsilon]
[ s\beta(1 + \alpha) - k^2_{\perp} + \imath \epsilon] [s\alpha \beta - k^2_{\perp} + \imath \epsilon]}
F_0\left(\frac{s}{ k^2_{\perp}}\right). 
\end{eqnarray}  
  
Applying the Cauchy theorem to integration over $\alpha$, we close up the integration contour and take residue in 
the pole 
  
    \begin{equation}\label{polealpha}
 \alpha = \frac{k^2_{\perp}}{1 - \beta} + \imath \epsilon. 
  \end{equation}

This pole is in the lower semi-plane when $0 < \beta <1$ whilst the poles in $\alpha$ generated by the other factors are in the 
upper semi-plane. So, closing the contour down and calculating the residue, we obtain the IREE as follows: 

\begin{eqnarray}\label{eqf0beta}
F_0\left(\frac{s}{\mu^2}\right) &=& 1 - \frac{\alpha_s C_F}{2 \pi} 
\int_{D_0} \frac{d k^2_{\perp}}{k^2_{\perp}} 
\frac{d \beta}{(\beta - k^2_{\perp}/s)} \;F_0\left(\frac{s}{k^2_{\perp}}\right). 
\end{eqnarray}

Eq.~(\ref{eqf0beta}) reads that integration over $\beta$ yields a logarithm in the region $\beta \gg k^2_{\perp}/s$. 
Requirements 

\begin{eqnarray}\label{d0}
0 &<& \beta < 1,
\\ \nonumber
\beta &\gg& k^2_{\perp}/s
\end{eqnarray}
define the integration  region $D_0$ in Eq.~(\ref{eqf0beta}). 
$D_0$ is depicted on graph (a) in Fig.~3. 
\begin{figure}\label{sud70fig3}
\includegraphics[width=.6\textwidth]{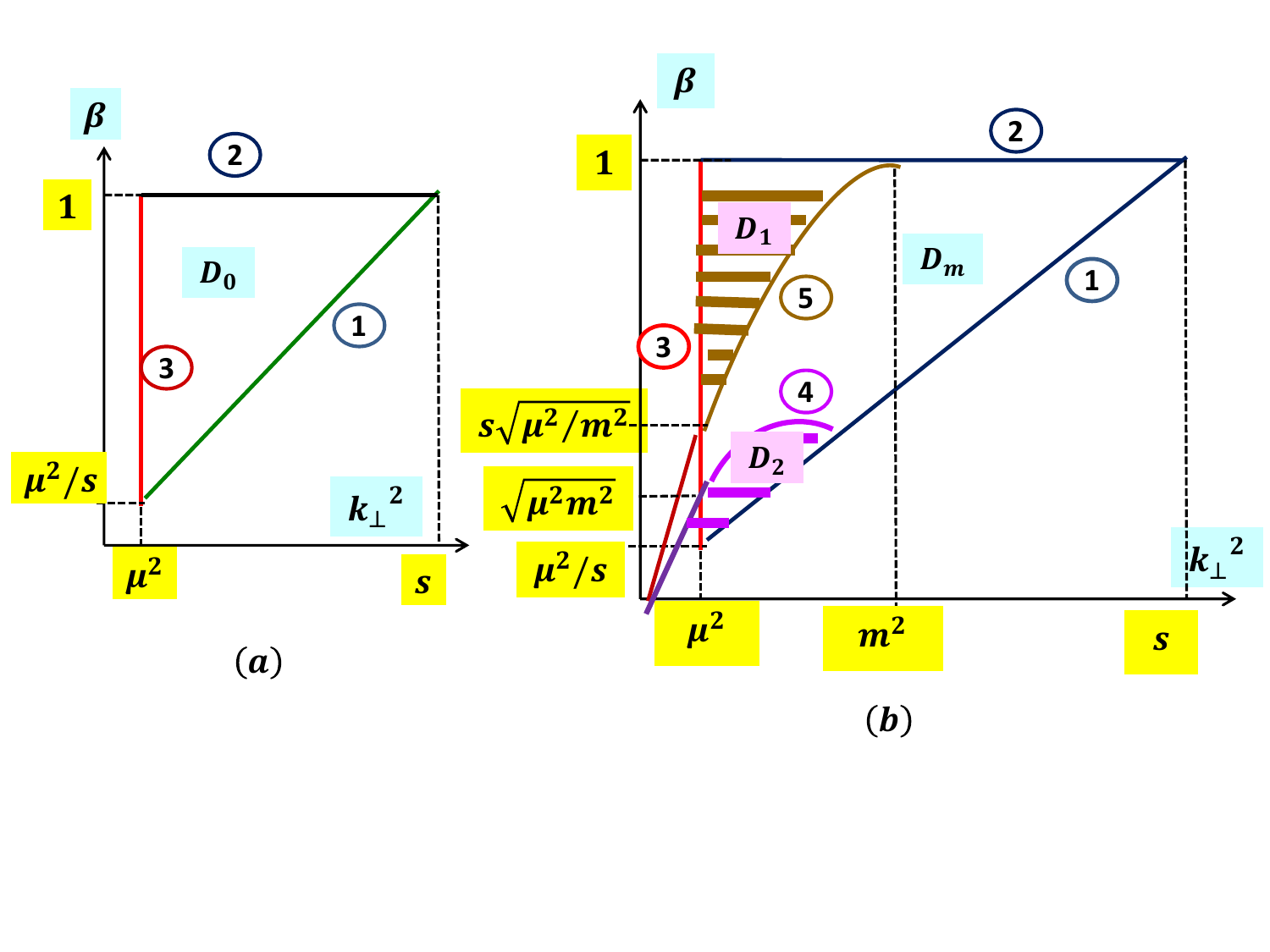}
\caption{\label{sud70fig3}
Integration region for the form factors $F_0$ (graph (a)) and $F_m$ (graph (b)) }
\end{figure}
Integration 
over $\beta$ does not involve $f(s/k^2_{\perp})$. As a result, we obtain  

\begin{eqnarray}\label{eqf0k}
F_0\left(\frac{s}{\mu^2}\right) &=& 1 - \frac{\alpha_s C_F}{2 \pi} \int_{\mu^2}^{s} \frac{d k^2_{\perp}}{k^2_{\perp}} 
\ln \left(\frac{s}{k^2_{\perp}}\right) \;F_0\left(\frac{s}{k^2_{\perp}}\right). 
\end{eqnarray}

Differentiating Eq.~(\ref{eqf0k}) with respect to $\ln (s/\mu^2)$, easily obtain the obvious solution 
   
\begin{equation}\label{f0w}
F_0 \left(\frac{s}{\mu^2}\right) = \exp \left[- W^{(1)}_{QCD}\right] =  \exp \left[- \frac{\alpha_s C_F}{4 \pi}\ln^2 \left(\frac{s}{\mu^2}\right)  \right]. 
\end{equation}

Eq.~(\ref{f0w}) reproduces the results of Refs.~\cite{quinn,smilga,tikt} but its derivation with using the IREE approach 
is much shorter then the derivations suggested in Refs.~\cite{quinn,smilga,tikt}.\\

\subsection{Accounting for the quark masses}

Let us account now for the quark masses impact on the form factor. 
Dependence on them appears when the IR cut-off $\mu$ is small: $\mu \ll m$. First,  
we drop $p^2_{1,2}$ in  Eq.~(\ref{eqf0gen}) and retain $m^2$. 
Then, we have to use a more involved parametrization than the one  in Eq.~(\ref{sudpar}). Namely, we have to introduce 
massless momenta $P_{1,2}$ which are made of $p_{1,2}$: 

\begin{eqnarray}\label{paron1}
P_1 &=& p_1 - y p_2,
\\ \nonumber
P_2 &=& p_2 - y p_1,
\end{eqnarray}
where 

\begin{equation}\label{y}
y = m^2/s,
\end{equation}
and represent $k$ as follows: 

\begin{equation}\label{paron}
k = \alpha P_2 + \beta P_1 + k_{\perp}.
\end{equation}

The transform inverse to (\ref{paron}) is 

\begin{eqnarray}\label{invparon}
p_1 &=& P_1 +  y P_2,
\\ \nonumber
p_2 &=& P_2 + y P_1,
\end{eqnarray}
so

\begin{eqnarray}\label{invsudon}
2 kp_1 &=& 2kP_1 +  y 2kP_2 = s \alpha + s y \beta,
\\ \nonumber
2kp_2 &=& 2kP_2 + y 2kP_1 = s \beta + s y \alpha,
\\ \nonumber 
k^2 &=& s \alpha \beta - k^2_{\perp}
\end{eqnarray}
and therefore Eq.~(\ref{eqf0gen}) is replaced by

  \begin{eqnarray}\label{eqfmgen}
F_m\left(\frac{s}{\mu^2},\frac{m^2}{\mu^2}\right)  
&=& 1 
-\imath \frac{\alpha_s C_F}{4 \pi^2} \int  \frac{d \alpha d \beta d k^2_{\perp} }
{R^{\prime}_1 (s \alpha, s \beta, k^2_{\perp}) R^{\prime}_2 (s \alpha, s \beta, k^2_{\perp}), R^{\prime}_3 (s \alpha, s \beta, k^2_{\perp})}
F_m\left(\frac{s}{ k^2_{\perp}},\frac{m^2}{ k^2_{\perp}}\right),
\end{eqnarray}
where $R^{\prime}_i = R_i|_{p^2_{1,2} = 0} $ and $R_i$ are defined in Eq.~(\ref{r123gen}). 
Then we use the Cauchy 
theorem for integration over $\alpha$ but in contrast to Eq.~(\ref{polealpha}) the pole is located now at 

\begin{equation}\label{apoleon}
\alpha  \approx -y \beta  - k^2_{\perp}. 
\end{equation}

Combining Eqs.~(\ref{invsudon}) and (\ref{apoleon}), we bring the second and third factors, $R^{\prime}_2$ and $R^{\prime}_3$  in  
Eq.~(\ref{eqfmgen}) to the following form: 

\begin{eqnarray}\label{ron23}
R^{\prime}_2 &=& s \beta + ys \alpha + (s \alpha \beta - k^2_{\perp}) = s \beta + s y \alpha - k^2_{\perp}  ,
\\ \nonumber
R^{\prime}_3  &=& s \alpha \beta - k^2_{\perp}  -  \beta (y \beta +  k^2_{\perp}) - k^2_{\perp} \approx -  y \beta^2 +  k^2_{\perp} .
\end{eqnarray}

$R^{\prime}_3$ can yield a logarithm when $y \beta^2 \ll k^2_{\perp}/s$, i.e. when

\begin{equation}\label{line5}
  \beta \ll \sqrt{k^2_{\perp}/sy}. 
\end{equation}

Then, integration of  $R^{\prime}_2$ over $\beta$ yields logarithm 
when $s \beta \gg k^2_{\perp}$ and $\beta \gg y \alpha$. Confronting the latter condition to Eq.~(\ref{line5}), obtain 
restrictions on $\beta$: 

\begin{equation}\label{line45}
\sqrt{k^2_{\perp}/sy} \gg \beta \gg \sqrt{y k^2_{\perp}/s}. 
\end{equation}

Note that $\sqrt{y k^2_{\perp}} > k^2_{\perp}$ when $k^2_{\perp} < m^2$. On the contrary, $\sqrt{y k^2_{\perp}} < k^2_{\perp}$   
at $k^2_{\perp} > m^2$. Therefore, Eq.~(\ref{line45}) holds at $k^2_{\perp} < m^2$ while it is replaced by  

\begin{equation}\label{line4}
\sqrt{k^2_{\perp}/sy} \gg \beta \gg  k^2_{\perp} 
\end{equation}
at $k^2_{\perp} < m^2$. The inequalities above form the integration region $D_m$ depicted on graph (b) in Fig.~3. 
The numbers in the circles in Fig.3 (b)  mark the following lines: 

\begin{eqnarray}\label{lines}
(1)&:& \beta = k^2_{\perp}/s,
\\ \nonumber
(2)&:& \beta = 1,
\\ \nonumber
(3)&:& k^2_{\perp} = \mu^2, 
\\ \nonumber 
(4)&:&  \beta = \sqrt{s y k^2_{\perp}}
\\ \nonumber
(5)&:&  \beta = \sqrt{k^2_{\perp}/y s}. 
\end{eqnarray} 

So, the region $D_m$ is defined and the IREE for the form factor $F_m \left(s/\mu^2,m^2/\mu^2\right)$ takes the 
following form:

\begin{eqnarray}\label{eqfmbeta}
F_m\left(\frac{s}{\mu^2},\frac{m^2}{\mu^2}\right) &=& 1 - \frac{\alpha_s C_F}{2 \pi} 
\int_{D_m} \frac{d k^2_{\perp}}{k^2_{\perp}} 
\frac{d \beta}{\beta} \;
F_m\left(\frac{s}{k^2_{\perp}},\frac{m^2}{k^2_{\perp}}\right).
\end{eqnarray} 

Integration over $\beta$ does not involve dealing with $F$, so it can be done as in the first-loop expressions. To this end, let us 
note that region $D_m$ can be represented as follow (see Fig.~3b): 

\begin{equation}\label{don}
D_m = D_0 - D_1 -D_2,
\end{equation}
where region $D_0$ is depicted on graph $(a)$ in Fig.~3. So, the integral in Eq.~(\ref{eqfmbeta}) can be written as follows: 

\begin{equation}\label{dd12}
\int_{D_{m}} \frac{d \beta d k^2_{\perp}}{ \beta \;k^2_{\perp}} = \int_{D_0}\frac{d \beta d k^2_{\perp}} 
{ \beta \;k^2_{\perp}} - \int_{D_{1}}\frac{d \beta d k^2_{\perp}}{ \beta \;k^2_{\perp}}
 - \int_{D_{2}}\frac{d \beta d k^2_{\perp}}{ \beta \;k^2_{\perp}},
\end{equation}
where (cf. Eq.~(\ref{eqf0k}))

\begin{equation}\label{intd0}
\int_{D_0} \frac{d \beta d k^2_{\perp}}{ \beta \;k^2_{\perp}} = \int_{\mu^2}^{s} \frac{d k^2_{\perp}}{k^2_{\perp}}
\int^1_{k^2_{\perp}/s} \frac{d \beta}{\beta}= 
\int_{\mu^2}^{s} \frac{d k^2_{\perp}}{k^2_{\perp}} \ln (s/k^2_{\perp}) 
\end{equation}
and 

\begin{equation}\label{intd1}
\int_{D_1}  \frac{d \beta d k^2_{\perp}} 
{ \beta \;k^2_{\perp}} = \frac{1}{2}\int^{m^2}_{\mu^2}\frac{d k^2_{\perp} }{k^2_{\perp} }\;\int^1_{k^2_{\perp}/m^2} \frac{d \beta^2}{\beta^2}
= \frac{1}{2}\int^{m^2}_{\mu^2}\frac{d k^2_{\perp} }{k^2_{\perp}} \ln (m^2/k^2_{\perp}),
\end{equation}

\begin{equation}\label{intd2}
\int_{D_2}  \frac{d \beta d k^2_{\perp}} 
{ \beta \;k^2_{\perp}} = \frac{1}{2}\int^{m^2}_{\mu^2}\frac{d k^2_{\perp} }{k^2_{\perp} }\;\int^1_{k^2_{\perp}/m^2} \frac{d \beta^2}{\beta^2}
= \frac{1}{2}\int^{m^2}_{\mu^2}\frac{d k^2_{\perp} }{k^2_{\perp}} \ln (m^2/k^2_{\perp})
\end{equation}

Substituting Eqs.~(\ref{intd0},\ref{intd1},\ref{intd2}) in Eq.~(\ref{eqfmbeta}) and solving it
\footnote{We skip technical details of solving that equation because we are going to 
present them in the next Sect. when consider off-shell quarks.}, conclude that 

\begin{equation}\label{f0wm}
F_m \left(\frac{s}{\mu^2},\frac{m^2}{\mu^2}\right) = \exp \left[- W^{(1)}_{m}\right] =  
\exp \left[- \frac{\alpha_s C_F}{4 \pi}\left[\ln^2 \left(\frac{s}{\mu^2}\right) - \ln^2 \left(\frac{m^2}{\mu^2}\right)  \right]\right]. 
\end{equation}

Note that if the quarks have different masses, $ W^{(1)}_{m}$ is given by the following expression: 

\begin{equation}\label{gamma1on}
W^{(1)}_{m} =  \frac{\alpha_s C_F}{4 \pi}\left[ \ln^2  \left(\frac{s}{\mu^2}\right) - \frac{1}{2} \ln^2 \left(\frac{m^2_1}{\mu^2}\right) - \frac{1}{2} \ln^2 
\left(\frac{m^2_2}{\mu^2}\right) \right].
\end{equation}

\section{Sudakov form factor with off-shell quarks}

Now let the quarks in Figs.~1,2 be off-shell:  $p^2_{1,2} \gg m^2$. Then terms $m^2$ in Eq.~(\ref{eqfqcdgen}) can be dropped. 
If IR cut-off $\mu$ is chosen so high that $\mu^2 \gg p^2_{1,2}$, the result coincides with $F_0$ and it 
strongly depends on $\mu$. 
We will demonstrate below that the $\mu$-dependence in the opposite case 
$\mu^2 \gg p^2_{1,2}$ is lesser than for $F_0$ and even can disappear at all. Below we consider the following cases:\\
 \textbf{(A)} One quark is on-shell while another quark is off-shell,\\
  \textbf{(B)} Both quarks are off-shell.\\
 Throughout this Sect. we will use the logarithmic variables  
$\rho, z_1, z_2$ defined as follows: 

\begin{equation}\label{pho}
\rho = \ln (s/\mu^2),~y_1 = \ln (p^2_1/\mu^2),~y_2 = \ln (p^2_2/\mu^2). 
\end{equation}

We assume that $p^2_1 \gg p^2_2$ and lift this restriction in final 
formulae. The Sudakov parametrization for off-shell quarks looks as follows: 

\begin{equation}\label{sudparoff}
k = \alpha p^{\prime}_2 + \beta p^{\prime}_1 + k_{\perp}, 
\end{equation}  
with $p^{\prime}_{1,2}$ being made of $p_{1,2}$: 

\begin{eqnarray}\label{pprimeoff}
p^{\prime}_1 &=& p_1 - x_1 p_2,
\\ \nonumber
p^{\prime}_2 &=& p_2 - x_2 p_1,
\end{eqnarray}
where 
\begin{equation}\label{x12}
x_1 = |p^2_1|/s,~~x_2 = |p^2_2|/s
\end{equation}
while $s = (p_1 + p_2)^2 \approx 2 p_1p_2$. Momenta $p^{\prime}_{1,2}$ are massless: $p^{\prime~2}_1 = p^{\prime~2}_2 \approx 0$. 
The inverse transform looks as follows: 

\begin{eqnarray}\label{invpprimeoff}
p_1 &=& p^{\prime}_1 +  x_1 p^{\prime}_2,
\\ \nonumber
p_2 &=& p^{\prime}_2 + x_2 p^{\prime}_1
\end{eqnarray}
and therefore 

\begin{eqnarray}\label{denomeoff}
R_1^{\prime \prime} &=& p^2_1 - s \alpha - s x_1 \beta + k^2 \approx s x_1 - s \alpha - k^2_{\perp},
\\ \nonumber
R_2^{\prime \prime}&=& p^2_2 + s \beta + s x_2 \alpha + k^2 \approx s x_2 + s \beta - k^2_{\perp},
\\ \nonumber 
R_3^{\prime \prime} &=& s \alpha \beta - k^2_{\perp},
\end{eqnarray}
where $R^{\prime \prime}_i = R_i|_{m^2 = 0} $ and $R_i$ are defined in Eq.~(\ref{r123gen}). Once more we make use of Cauchy 
theorem when integrate over $\alpha$, closing the contour upwards and taking residue at the pole 

\begin{equation}\label{pole}
\alpha = x_1 - k^2_{\perp}/s + 
 \imath \epsilon,
\end{equation}
where 

\begin{eqnarray}\label{r23off}
R_2^{\prime \prime}
&=&  s \beta + s x_2 + x_1 \beta- k^2_{\perp}, 
\\ \nonumber
R_3^{\prime \prime} &=& x_1 \beta- k^2_{\perp}. 
\end{eqnarray}

Our aim is to find the integration region $D_4$, where $R_2^{\prime \prime} \approx s \beta$ and $R_3^{\prime \prime} \approx k^2_{\perp}$. 
 It follows from Eq.~(ref{r23off}), that the borders of $D_4$ are in the first place formed the following lines 
 (we number the lines in accordance with Eq.~(\ref{lines})): 
 
 \begin{eqnarray}\label{doff}
\textbf{(3):} &~& k^2_{\perp} \gg \mu^2,
\\ \nonumber 
\textbf{(6):} & ~& \beta \gg  x_2 + k^2/s,
\\ \nonumber
\textbf{(7):} &~& \beta \ll  k^2_{\perp}/ sx_1. 
\end{eqnarray}

However, these lines do not fix $D_4$ unambiguously.  Actually, the integration region strongly depends on the 
relations between the IR cut-off and quark virtualities, dividing $D_4$ into regions $D_{MV}$ and $D_{DV}$ as shown in Fig.~4. 
\begin{figure}\label{sud70fig4}
\includegraphics[width=.6\textwidth]{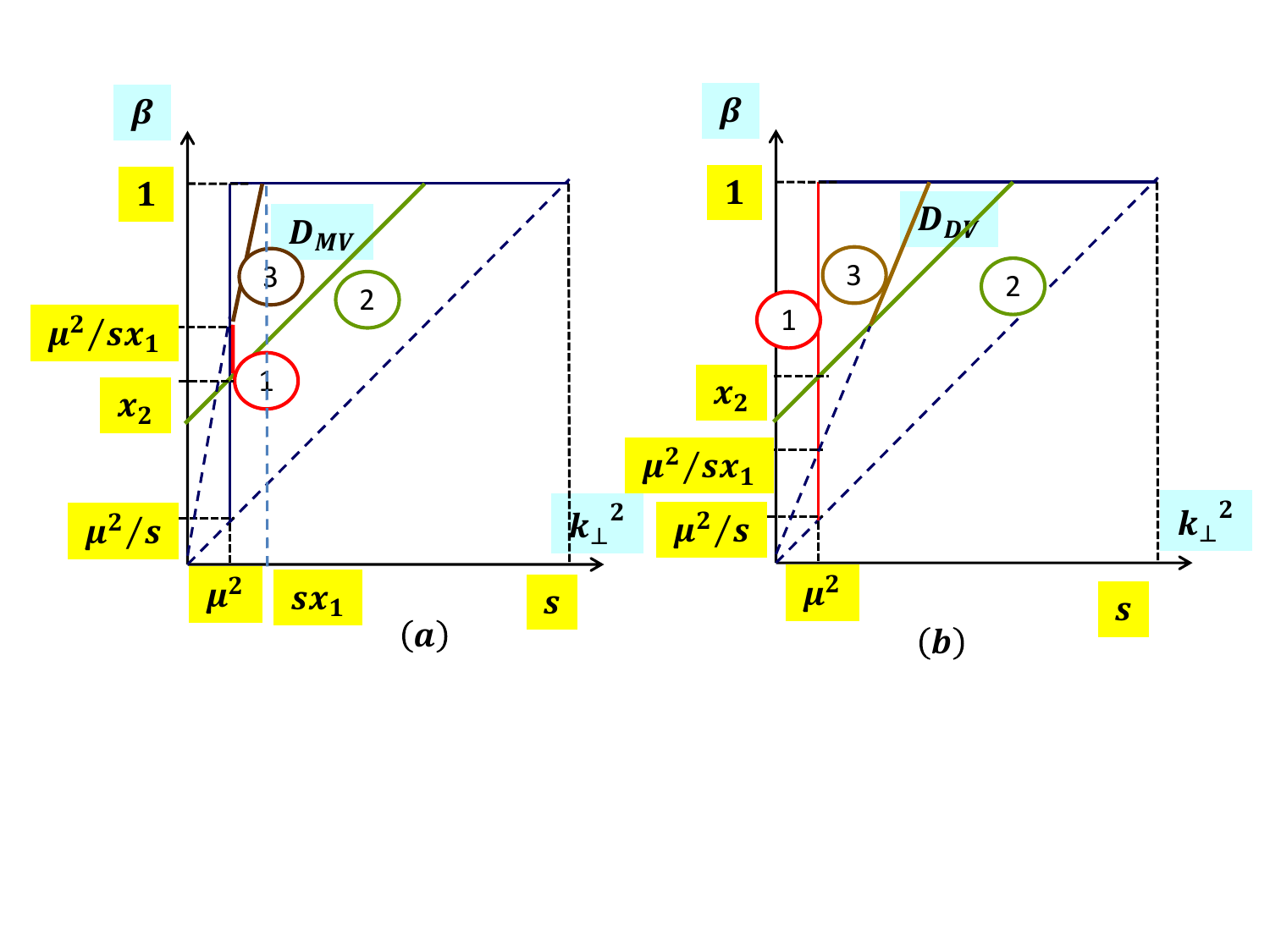}
\caption{\label{sud70fig4}
Moderately-virtual (Fig.~(a)) and deeply-virtual (Fig.~(b)) kinematic regions }
\end{figure}
First, there is the moderately-virtual(MV) kinematics, where 
the quark virtualities 
obey the following inequality: 

\begin{equation}\label{mvkin}
s \mu^2 \gg p^2_1 p^2_2.
\end{equation}

Region $D_{MV}$ is depicted on graph (a) in Fig.~4. 
Then, there is the deeply-virtual (DV) kinematics depicted on graph (b), where  

\begin{equation}\label{dvkin}
s \mu^2 \ll p^2_1 p^2_2. 
\end{equation}
 
This region does not involve line 1, so it is IR-stable. Before 
dealing with them, 
we start with considering a simpler situation, where one of the quarks is on-shell whereas the other is off-shell. We 
name such situation partly off-shell. 

\subsection{Partly off-shell form factor}

To be specific, let us consider the case when $p^2_1 \gg m^2$ whereas $p^2_2 \sim m^2$, i.e. $x_2 = 0$.  In this case the 
form factor $F_1 = F_1 (s/\mu^2, p^2_1/\mu^2)$. In terms of the logarithmic variables, $F_1 = F_1 (\rho, x_1)$.  
 The integration region $\widetilde{D}_1$ over the factorized gluon momentum is restricted by lines 1,6,7 of Eq.~(\ref{doff}), 
 however with $x_2 = 0$, and depicted in Fig.~5.
\begin{figure}\label{sud70fig5}
\includegraphics[width=.6\textwidth]{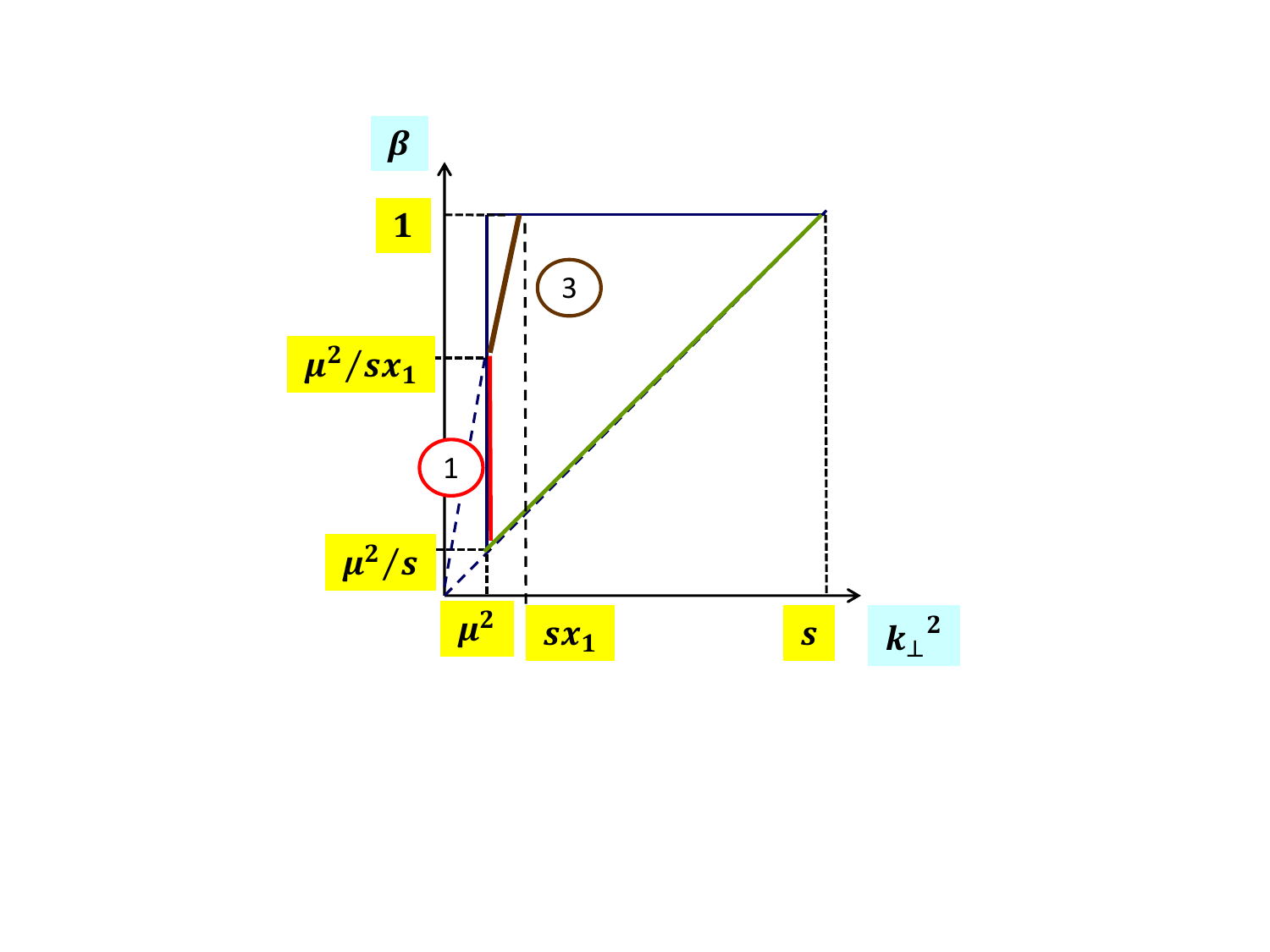}
\caption{\label{sud70fig5}
Integration region at $x_2 = 0$ }
\end{figure}
  The IREE for $F_1$ is 
 
 \begin{eqnarray}\label{inteqfx1gen}
F_1 \left(\frac{s}{\mu^2}, \frac{p^2_1}{\mu^2}\right) = 1 
 - 2\bar{a}\; 
 \left[\int_{\mu^2}^s \frac{d k^2_{\perp}}{k^2_{\perp}}\int^1_{k^2_{\perp}/s} \frac{d \beta}{\beta}
- \int_{\mu^2}^{p^2_1} \frac{d k^2_{\perp}}{k^2_{\perp}}\int^1_{k^2_{\perp}/sx_1} \frac{d \beta}{\beta}\right]
F_1 \left(\frac{s}{k^2_{\perp}}, \frac{p^2_1}{k^2_{\perp}}\right)
,
\end{eqnarray}
with $\bar{a} = \alpha_s C_F/4 \pi$. Integration over $\beta$ does not involve $F_1$, so after integration obtain:

\begin{eqnarray}\label{inteqfx1}
F_1 \left(\frac{s}{\mu^2}, \frac{p^2_1}{\mu^2}\right) = 1
 - 2\bar{a}\; 
 \left[\int_{\mu^2}^s \frac{d k^2_{\perp}}{k^2_{\perp}}\ln \left(\frac{s}{k^2_{\perp}}\right)
- \int_{\mu^2}^{p^2_1} \frac{d k^2_{\perp}}{k^2_{\perp}}
\ln \left(\frac{p^2_1}{k^2_{\perp}}\right)\right]
F_1 \left(\frac{s}{k^2_{\perp}}, \frac{p^2_1}{k^2_{\perp}}\right). 
\end{eqnarray}

Applying operator $-\mu^2 d /d \mu^2$ to Eq.~(\ref{inteqfx1}), obtain the differential equation: 

\begin{eqnarray}\label{eqfx1}
\left[\frac{\partial  }{\partial \rho} + \frac{\partial  }{\partial y_1} \right] 
F_1 \left(\rho,y_1\right) = 
 - 2 \bar{a}\; 
 \left(\rho
- y_1\right)
F_1 \left(\rho,y_1\right). 
\end{eqnarray}

In order to simplify Eq.~(\ref{eqfx1}) introduce new variables $z_{1,2}$ instead of $\rho,y_1$: 

\begin{eqnarray}\label{z12}
z_1 &=& \rho + y_1, 
\\ \nonumber
z_2 &=& \rho - y_1.
\end{eqnarray}

The inverse transform is 

\begin{eqnarray}\label{z12inv}
\rho &=& \frac{1}{2} [z_1 + z_2], 
\\ \nonumber
y_1 &=& \frac{1}{2} [z_1 - z_2].
\end{eqnarray}

Eq.~(\ref{eqfx1}) takes the following form: 

\begin{eqnarray}\label{eqfx1z12}
\frac{\partial  }{\partial z_1} 
F_1 \left(z_1,z_2\right) = 
 - \bar{a}\;  
z_2
F_1 \left(z_1,z_2\right), 
\end{eqnarray}

The general solution to it: 

\begin{equation}\label{fx1gen}
F_1 \left(z_1,z_2\right) = \Phi (z_2) e^{- \bar{a}z_1 z_2},
\end{equation}
where $\Phi (z_2)$ is arbitrary. To specify it we use the matching with the on-shell form factor, where quark masses are neglected: 

\begin{equation}\label{matchz2}
F_1 \left(z_1,z_2\right)|_{y_1 = 0} = F_0(\rho) = e^{- \bar{a}\rho^2}.
\end{equation}

Thus we obtain 

\begin{equation}\label{phi}
\Phi (\rho) e^{- \bar{a}\rho^2} = e^{- \bar{a}\rho^2},
\end{equation}
i.e. $\Phi = 1$. Hence, the partly off-shell form factor is given by the following expression: 

\begin{equation}\label{fy1}
F_1 \left(\frac{s}{\mu^2}, \frac{p^2_1}{\mu^2}\right) = \exp \left[- \frac{\alpha_s C_F}{4 \pi}\left[
\ln^2 \left(\frac{s}{\mu^2}\right) - \ln^2 \left(\frac{p^2_1}{\mu^2}\right) \right] \right].
\end{equation}

\subsection{Off-shell form factor in MV kinematic region}

Consider the form factor $F_{MV} = F_{MV}(s,p^2_1,p^2_2)$ in the case when the both quarks are off-shell and their  
virtualities $p^2_1,p^2_2$ obey Eq.~(\ref{mvkin}). IREE for $F_{MV}$ is 

\begin{equation}\label{eqfmv}
\frac{\partial F_{MV}}{\partial \rho} + \frac{\partial F_{MV}}{\partial y_1} + \frac{\partial F_{MV}}{\partial y_2} = 
- 2 \bar{a} (\rho - y_1 - y_2) F
\end{equation}
and the general solution to it is

\begin{equation}\label{fmvgen}
F_{MV}^{gen} (\rho, y_1, y_2)= \Phi_{MV} (\rho - y_1, \rho - y_2) e^{- \bar{a}\left[\rho^2 - y^2_1 - y^2_2\right]}.
\end{equation}

Specify it with the matching 

\begin{equation}\label{matchy2}
F_{MV}^{gen} (\rho, y_1, y_2)|_{y_2 = 0} = F_1(\rho.y_1). 
\end{equation}
It yields $\Phi_{MV}  = 1$, so 

\begin{equation}\label{fmv}
F_{MV} (\rho, y_1, y_2) = e^{- \bar{a}\left[\rho^2 - y^2_1 - y^2_2\right]}. 
\end{equation}

\subsection{Off-shell form factor in DV kinematic region}

Fig.~4b shows that DV region does not involve the cut-off $\mu$, so the form factor $F_{DV}$ DV region does not 
depend on $\mu$ and therefore IREE for $F_{DV}$ is

\begin{equation}\label{eqfdv}
\frac{\partial F_{DV}}{\partial \rho} + \frac{\partial F_{DV}}{\partial y_1} + \frac{\partial F_{DV}}{\partial y_2} = 0
\end{equation}
and the general solution to it is

\begin{equation}\label{fdvgen}
F_{DV}^{gen} (\rho, y_1, y_2)= \Phi_{DV} (\rho - y_1, \rho - y_2), 
\end{equation}
with $\Phi_{DV}$ being arbitrary. To specify it, we use the matching: 

\begin{equation}\label{matchy3}
 F_{DV} (\rho, y_1, y_2) = F_{MV} (\rho, y_1, y_2) 
\end{equation}
at $\rho = y_1 + y_2$. It means that 

\begin{equation}\label{phidv}
\Phi_{DV} ((y_2,y_1) = e^{- \bar{a}\left[(y_1 + y_2)^2 - y^2_1 - y^2_2\right]} = e^{- 2\bar{a}y_1 y_2}. 
\end{equation}

Replacing $y_1$ ($y_2$) with $\rho - y_2$ ($\rho - y_1$), we obtain 

\begin{equation}\label{matchy4}
F_{DV} (\rho, y_1, y_2) = e^{- 2\bar{a}(\rho -y_1)(\rho - y_2)} = \exp \left[- \frac{\alpha_s C_F}{2 \pi}\ln(s/p^2_1)\ln (s/p^2_2)\right]. 
\end{equation}

Obviously, $F_{DV}$ is IR-stable. To conclude, we remind that direct transition from $F_{DV} $ to form factors with on-shell quarks 
brings incorrect results.  To avoid it, one needs first to move from $F_{DV}$ to $F_{MV}$ and then  put $p^2_{1,2} = \mu^2$.

\section{Sudakov form factor for electroweak reactions} 

Sudakov form factor $F_{EW}$ for electroweak (EW) high-energy reactions was considered in DLA in Ref.~\cite{flmm} by constructing and solving 
IREE, though some essential technical details were skipped in that article. Below we fill in this gap by presenting a more consistent and 
detailed way to calculate $F_{EW}$, making use of the results obtained in the previous Sects. Besides, we account for impact on 
$F_{EW}$ of masses of fermions (quarks or leptons) involved. Main difference between form factors $F_{EW}$ and $F$ in QCD/QED is 
that, in addition to the massless i.e. virtual photon exchanges, exchanges with $W,Z$ bosons have to be also accounted for. 
We do not account for the mass difference of $W$ and $Z$ -bosons and use the same notation $M$ for their masses. First we assume 
that masses of the fermions involved are negligibly small and than account for the impact o the masses on  
$F_{EW}$.

\subsection{massless fermions}

In this Sect.~ we re-derive the results obtained in Ref.~\cite{flmm}, presenting them in a more detailed way. Obviously,  
virtual EW boson exchanges are IR-stable because $M$ acts as an IR cut-off, regulating IR divergences 
involving $W,Z$ bosons. On the contrary, regulating IR divergences in photon exchanges 
requires introducing an IR cut-off $\mu$. There can be two situations: \\
\textbf{(i)} $\mu \geq M$. In this case, we denote the form factor $\widetilde{F}_{EW}$  
and keep the IR cut-off $ = M$, 
so 
$\widetilde{F}_{EW} = \widetilde{F}_{EW}(s/M^2)$.
\\
\textbf{(ii)} $\mu \ll M$. In this case, $F_{EW} = F_{EW} (s/\mu^2, M^2/\mu^2)$, i.e. $F_{EW}$ contains logs of 
both $s/\mu^2$ and $M^2/\mu^2$. 

Consider first situation \textbf{(i)}. IREE for $\widetilde{F}_{EW}$ is pretty similar to the one for $F_0$, see 
Eq.~(\ref{eqf0k}), and the integration region coincides with the one in Fig.~3a providing that $\mu$ is replaced by $M$: 

\begin{equation}\label{eqftildeew}
\widetilde{F} \left(\frac{s}{M^2}\right) = 1  
 - (c_{\gamma} + c_{EW}) \int^s_{M^2} \frac{d k^2_{\bot}}{k^2_{\bot}} \ln \left(\frac{s}{k^2_{\bot}}\right) 
\widetilde{F} \left(\frac{s}{k^2_{\bot}}\right), 
\end{equation}
where

\begin{eqnarray}\label{cew}
c_{\gamma} 
&=& \frac{\alpha}{2 \pi} Q^2,
\\ \nonumber
c_{EW} + c_{\gamma}&=&  \frac{1}{ 8 \pi^2} \left[g^2 \tau_a \tau_a + g^{\prime 2} \frac{Y^2}{4}  \right] , 
\end{eqnarray}
where all notations in the r.h.ss. are standard: 
$Q$ stands for the fractional electric charge of the fermions, $\tau_a$ and $Y$ are generators of $SU(2)$ and $U(1)$ groups respectively, 
notations $g$ and $g^{\prime}$ 
are for the fermion-boson $SU(2)$ and $U(1)$ couplings respectively.
Solution to Eq.~(\ref{eqftildeew}) is

\begin{equation}\label{ftildeew1}
\widetilde{F} \left(\frac{s}{M^2}\right) = \exp \left[-\frac{1}{2}(c_{\gamma} + c_{EW}) \ln^2 
\left(\frac{s}{M^2}\right)\right]. 
\end{equation}

Obviously, $M$ acts in Eq.~(\ref{ftildeew1}) as a universal IR cut-off for all electroweak bosons. \\
Now consider situation \textbf{(ii)}:  
IREE for 
$F_{EW}$ is more complicated than Eqs.~(\ref{eqf0k},\ref{eqftildeew}), although its pattern also corresponds 
to Fig.3a. Handling integration over $\beta$ is the same but integration over $k^2_{\bot}$ is more involved. Primarily, 
the integration region is 
$L:~~ \mu^2 \ll k^2_{\bot} \ll s$. 
We divide it in two parts, $L_{1,2}$: \\

\begin{eqnarray}\label{kregtot}
L_1:~~ M^2 \ll k^2_{\bot} \ll s, 
\\ \nonumber 
L_2:~~ \mu^2 \ll k^2_{\bot} \ll M^2.
\end{eqnarray}

Note that integration over $L_1$ is identical to the one in Eq.~(\ref{eqftildeew}), so it yields $\widetilde{F}$ which  
 becomes the first term in the r.h.s. of the IREE we are constructing. Then note that the remaining integration over $L_2$ 
can yield logs from factorized virtual photons only. It allows us to fix the second (integral) term in Eq.~(\ref{inteqfew}):

\begin{equation}\label{inteqfew}
F_{EW} \left(\frac{s}{\mu^2}, \frac{M^2}{\mu^2}\right) = \widetilde{F} \left(\frac{s}{M^2}\right)
 - c_{\gamma} \int^{M^2}_{\mu^2} \frac{d k^2_{\bot}}{k^2_{\bot}} \ln \left(\frac{s}{k^2_{\bot}}\right) 
F_{EW}\left(\frac{s}{k^2_{\bot}},\frac{M^2}{k^2_{\bot}}\right) .
\end{equation}

Applying operator $- \mu^2 d/ d \mu^2$ converts  Eq.~(\ref{inteqfew}) into a differential equation: 

\begin{eqnarray}\label{eqfew}
\frac{\partial F_{EW}}{\partial \rho} +  \frac{\partial F_{EW}}{\partial \eta} = 
c_{\gamma} (\rho - \eta) F_{EW},
\end{eqnarray}
where  $\rho = \ln (s/\mu^2)$ and  $\eta = \ln (M^2/\mu^2)$. Introducing variables $u,v$:

\begin{eqnarray}\label{uvdef}
u &=& \rho + \eta, 
\\ \nonumber
v &=& \rho - \eta,
\end{eqnarray}
we obtain 

\begin{eqnarray}\label{eqfuv}
\frac{\partial F_{EW}}{\partial u} = -\frac{1}{2}
c_{\gamma} v F_{EW}. 
\end{eqnarray}

A general solution to Eq.~(\ref{eqfuv}) is 

\begin{eqnarray}\label{fuvgen}
 F_{EW} = \Phi_{EW} (v) e^{-\frac{1}{2} c_{\gamma} u v F_{EW}},  
\end{eqnarray}
where $\Phi_{EW}$ is arbitrary. We fix it through the matching 

\begin{equation}\label{matchew}
F_{EW}|_{\mu = M} = \widetilde{F},
\end{equation}

i.e. 

\begin{equation}\label{matchewrho}
\Phi (\rho) e^{-\frac{1}{2} c_{\gamma} \rho^2 } = e^{-\frac{1}{2}(c_{\gamma} + c_{EW})\rho^2}
\end{equation}
and therefore 

\begin{equation}\label{phiv}
\Phi (v) =  e^{-\frac{1}{2} c_{EW} v^2 }
\end{equation}
Combining Eqs.~(\ref{phiv}) and (\ref{fuvgen})  leads to 

\begin{equation}\label{few}
F_{EW} \left(\frac{s}{\mu^2}, \frac{M^2}{\mu^2}\right) = \exp \left[- \frac{1}{2} \left[c_{\gamma} \left[\ln^2 
\left(\frac{s}{\mu^2}\right)
- \ln^2 \left(\frac{M^2}{\mu^2}\right)\right] + c_{EW} \ln^2 \left(\frac{s}{M^2}\right)\right]\right].
\end{equation}

\subsection{Massive fermions}

When the fermion mass $m$ obeys  

\begin{equation}\label{msmall}
\mu \ll m \ll M,
\end{equation}

\begin{equation}\label{fewmsmall}
F_{EW} \left(\frac{s}{\mu^2}, \frac{M^2}{\mu^2},\frac{m^2}{\mu^2}\right) = \exp \left[- \frac{1}{2} \left[c_{\gamma} \left[\ln^2 
\left(\frac{s}{\mu^2}\right)
- \ln^2 \left(\frac{M^2}{\mu^2}\right) - \ln^2 \left(\frac{m^2}{\mu^2}\right) \right] + c_{EW} \ln^2 \left(\frac{s}{M^2}\right)\right]\right].
\end{equation}

When $m$ is so great that

\begin{equation}\label{mbig}
 m \gg M,
\end{equation}

\begin{equation}\label{ftildeew}
\widetilde{F} \left(\frac{s}{M^2},\frac{m^2}{M^2}\right) = \exp \left[-\frac{1}{2}(c_{\gamma} + c_{EW}) \left[ \ln^2 
\left(\frac{s}{M^2}\right) - \ln^2 \left(\frac{m^2}{M^2}\right) \right]\right]. 
\end{equation}

\section{Inelastic Sudakov form factor}

In the present Sect. we demonstrate that using the IREE method makes it possible to easily calculate inelastic form factors in DLA. Consider 
amplitude  of photon production of quark-antiquark pair complementing emission of a single bremsstrahlung gluon: 
$\gamma^{\star} \to q (p_1)~\bar{q}(p_2) + g (k_1)$. This amplitude, $A_1 (p_1,p_2,k_1,l)$, with 
$l$ being the polarization vector of the gluon, is depicted in Fig.~6.
\begin{figure}\label{sud70fig6}
\includegraphics[width=.6\textwidth]{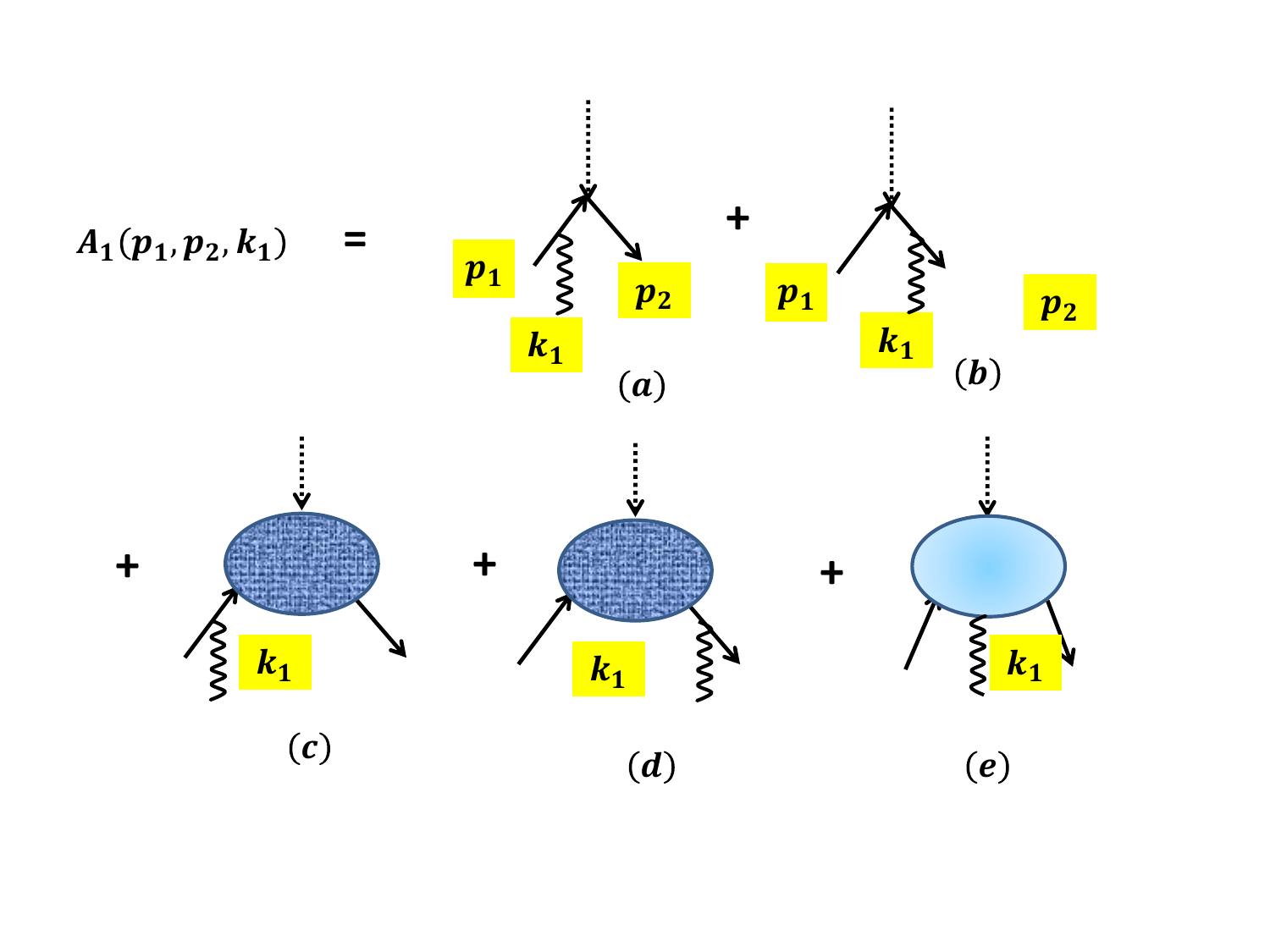}
\caption{\label{sud70fig6}
Graphs for amplitude of the single-gluon emission, with $k_1$ being the gluon momentum }
\end{figure}
 The quarks are on-shell and their masses are neglected.  Under certain conditions, $A_1$ takes the factorized form: 

\begin{equation}\label{f1def}
A_1 = B(k_1) f_1 (p_1,p_2,k_1), 
\end{equation}
where  $f_1 (p_1,p_2,k)$ is the quark inelastic form factor while 
$B(k_1)$ denotes the bremsstrahlung factor: 

\begin{equation}\label{bdef}
B(k_1) = \left(\frac{p_2 l}{p_2k_1} - \frac{p_1 l}{p_1k_1}\right).
\end{equation}

Primarily, a bremsstrahlung factorization condition was obtained in Refs.~\cite{low1,low2}. It is called 
Low-Gell-Mann-Goldberger (LGMG) theorem. This theorem  
refers to  emission of photons with very small energies $\omega$ and $F_1$ coincides with the elastic 
(non-radiative) form factor which  
does not depend on $k_1$. The key argument there 
was that $B(k_1) \sim 1/\omega$, so $B(k_1)$ is great at $\omega \to 0$. The theoretical tool of the proof is usage of the gauge 
invariance: $ A_1 (p_1,p_2,k_1,k_1)= 0$, when $l$ in $ A_1 (p_1,p_2,k_1,l)$, is replaced by $k_1$. It relates graphs 
(c,d,e) in Fig.~6 and makes it possible to express the inelastic blob (e) through off-shell elastic blobs (c) and (d). Then 
LGMG theorem was generalized to gluon emission  in Ref.~\cite{l}. Note that  LGMG theorem holds for both high and low energies. 

In the high-energy case, the bulk of bremsstrahlung is concentrated in independent cones located along the quark directions,  
where $B \sim 1/k_{1\bot}$. This observation was the starting point of Gribov bremsstrahlung theorem\cite{g} which 
was proved also in the QED context. This theorem 
states that in the case of photon emission at high energies the factor $B(k_1)$ can be factorized providing 
$k_{1\bot}$ is small though the photon energy $\omega$ does not have to be small. 
Generalization of this theorem to QCD was done in Ref.~\cite{ce}.  
Emission of one bremsstrahlung gluon with DL accuracy in the first loop was studied in Ref.~\cite{kf}, where the 
essential difference 
between amplitudes of this process in QED and QCD was demonstrated. 
Explicit expressions for  
inelastic form factors $f_n(p_1,p_2,k_1,..,k_n)$ (with the number of emitted gluons $n= 1,2,...$) were calculated in DLA in 
Ref.~\cite{efl} and confirmed later in Ref.~\cite{kurcher}. 
 
Below we explain how to calculate $f_n$, keeping, in contrast to in Ref.~\cite{efl}, the same way for constructing IREEs as in 
the previous Sects.  We start with constructing IREE for $F_1$ and depict it in Fig.~7. 
\begin{figure}\label{sud70fig7}
\includegraphics[width=.6\textwidth]{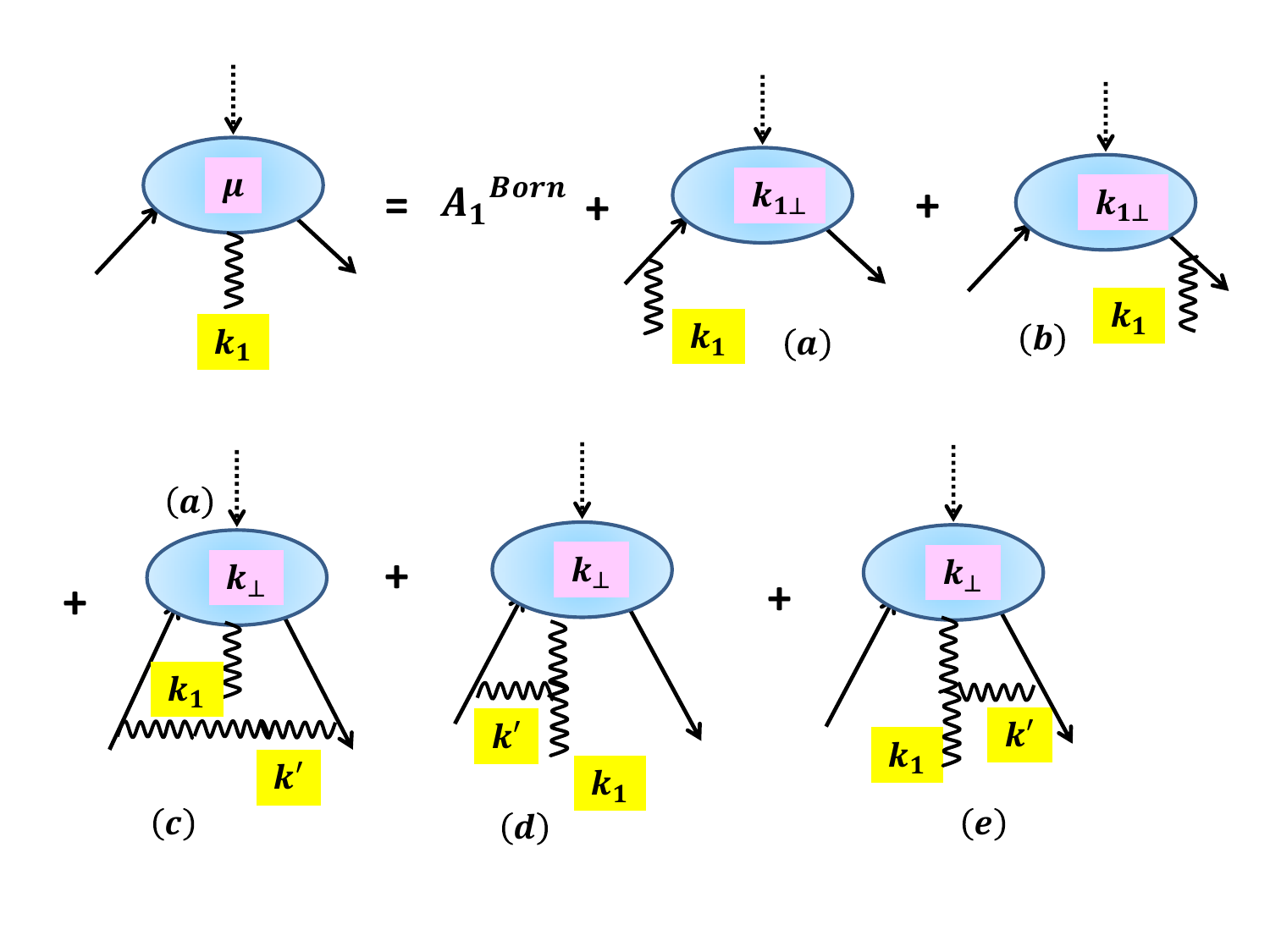}
\caption{\label{sud70fig7}
IREE for the inelastic form factor with emission of the single gluon }
\end{figure}
The l.h.s. of the IREE is $F_1$ as depicted in l.h.s. in Fi.~7. In order to construct the r.h.s. we need to introduce IR cut-off 
$\mu$  and look for gluons with minimal transverse momenta. 
Denote $k^{\prime}_i$ momenta of virtual gluons. 
In order to get logarithmic contributions from integrations over transverse momenta $k^{\prime}_{i \bot}$ , the 
moments should obey 
strong inequalities requirements: $\mu \ll k^{\prime}_{i \bot} \ll k^{\prime}_{j \bot}$ ($i \neq j, i,j = 1,2,...$). 
Then denote $k^{\prime}_{\bot} = \min \{k^{\prime}_{i \bot}\}$ the minimal transverse momentum of virtual gluons   
and compare it with transverse momentum $k_{1 \bot}$ of the emitted gluon. There can be two situations: \textbf{(i)}: 
$k_{1 \bot} \ll k^{\prime}_{ \bot}$ and \textbf{(ii)} $k^{\prime}_{ \bot} \ll k_{1 \bot}$. 
Consider them separately. When 

\begin{equation}\label{k1small}
\mu \ll k_{1 \bot} \ll k^{\prime}_{ \bot},
\end{equation}
the emitted gluon can be factorized out of $A_1$ as shown on graphs (a) and (b) in Fig.~7. 
The blobs on these graphs correspond to elastic form factor $F_0$ defined in Eq.~(\ref{f0w}), where cut-off $\mu$ 
is replaced by $k_{1 \bot}$. 
In the opposite case, when 

\begin{equation}\label{kprimesmall}
\mu \ll k^{\prime}_{ \bot} \ll k_{1 \bot},  
\end{equation}
the virtual gluon can be factorized as shown on graphs (c,d,e) whereas the blobs correspond to $f_1$, 
where $k^{\prime}_{ \bot}$ acts as IR cut-off 
for other virtual gluons. Note that graphs (d) and (e) are absent in QED. As all the blobs in Fi.~7 do not depend on 
longitudinal components of the factorized gluons, integration over them is carried out as in the first-loop case. 
 After that we arrive at the following IREE: 

\begin{eqnarray}\label{eqf1def}
f_1 \left(\frac{s}{\mu^2},\frac{k^2_{1 \bot}}{\mu^2} \right) &=& F_0\left(\frac{s}{k^2_{\bot}}\right) 
- \frac{\alpha_s}{2 \pi} \left(C_F - \frac{N}{2}\right) 
\int^{k^2_{1 \bot}}_{\mu^2} \frac{d k^{\prime 2}_{ \bot}}{k^{\prime 2}_{ \bot}} 
\ln \left(s/k^{\prime 2}_{ \bot}\right)
f_1 \left(\frac{s}{k^{\prime 2}_{ \bot}}, \frac{k^2_{1 \bot}}{k^{\prime 2}} \right)
\\ \nonumber
&-& \frac{\alpha_s}{2 \pi}\;\frac{N}{2} \left[
\int^{k^2_{1 \bot}}_{\mu^2} \frac{d k^{\prime 2}_{ \bot}}{k^{\prime 2}_{ \bot}} 
\ln \left(s_1/k^{\prime 2}_{ \bot}\right) f_1 \left(\frac{s}{k^{\prime 2}_{ \bot}}, \frac{k^2_{1 \bot}}{k^{\prime 2}} \right)
+ \int^{k^2_{1 \bot}}_{\mu^2} \frac{d k^{\prime 2}_{ \bot}}{k^{\prime 2}_{ \bot}} 
\ln \left(s_2/k^{\prime 2}_{ \bot}\right) f_1 \left(\frac{s}{k^{\prime 2}_{ \bot}}, \frac{k^2_{1 \bot}}{k^{\prime 2}} \right)\right],
\end{eqnarray}
where we have denoted $s_1 = 2p_1k, s_2 = 2p_2k$ and $F_0$ is defined in Eq.~(\ref{f0w}). Using  that $s_1 s_2 = s k^2_{1 \bot}$, we bring Eq.~(\ref{eqf1def}) to the 
following form: 

\begin{eqnarray}\label{eqf1w}
f_1 \left(\frac{s}{\mu^2},\frac{k^2_{1 \bot}}{\mu^2} \right) &=& F_0\left(\frac{s}{k^2_{\bot}}\right) 
- \frac{\alpha_s}{2 \pi}\;
 \int^{k^2_{1 \bot}}_{\mu^2} \frac{d k^{\prime 2}_{ \bot}}{k^{\prime 2}_{ \bot}} 
\left[C_F \ln \left(\frac{s}{ k^{\prime 2}_{ \bot}}\right) + \frac{N}{2} \ln \left(\frac{k^2_{1 \bot}}{ k^{\prime 2}_{ \bot}}\right) \right]
f_1 \left(\frac{s}{k^{\prime 2}_{ \bot}}, \frac{k^2_{1 \bot}}{k^{\prime 2}} \right).
\end{eqnarray}

When Eq.~(\ref{eqf1w}) is differentiated over $\mu$, the term $F_0\left(s/k^2_{\bot}\right)$ vanishes. It is convenient to solve  the obtained 
differential equation in terms of logarithmic variables $x, y$: $x = \ln (s/)\mu^2, y = \ln(k^2_{1 \bot}/\mu^2)$, arriving at 
the partial differential equation  

\begin{equation}\label{eqf1xy}
\frac{\partial f_1}{\partial x} + \frac{\partial f_1}{\partial y} = -\left[\lambda_1 x + \lambda_2 y\right] f, 
\end{equation}
where

\begin{equation}\label{lambda12}
\lambda_1 = \frac{\alpha_s}{2 \pi} C_F,
\\ \nonumber 
\lambda_2 = \frac{\alpha_s}{2 \pi}\;\frac{N}{2}.
\end{equation}

In turn, this equation can easily be reduced down to the ordinary equation 

\begin{equation}\label{eqf1uv}
\frac{\partial f_1}{\partial u}  = -\frac{1}{2}\;[\lambda_+ u + \lambda_- y] f,
\end{equation}
in terms of variables $u,v$:

\begin{eqnarray}\label{uvxydef}
u = x + y,~~v= x - y
\end{eqnarray}
and with $\lambda_{\pm} = \lambda_1 \pm \lambda_2$. Solving Eq.~(\ref{eqf1uv}) and specifying the general solution 

\begin{equation}\label{f1gen}
f_1 = \Phi (v) \exp \{- \frac{1}{8}\left[\lambda_+ u^2 + \lambda_- uv\right]\}
\end{equation}
with matching 

\begin{equation}\label{matchf1f0}
f_1(s/\mu^2, k^2_{1 \bot}/\mu^2)|_{k^2_{1 \bot} = \mu^2} = F_0 (s/\mu^2),
\end{equation}
with $F_0$ defined in Eq.~(\ref{f0w}), we obtain 

\begin{equation}\label{f1}
f_1 \left(\frac{s}{\mu^2}, \frac{k^2_{1 \bot}}{\mu^2}\right) = \exp \left\{- \frac{\alpha_s}{2 \pi}\left[C_F \ln^2\left(\frac{s}{\mu^2}\right) + 
\frac{N}{2}\ln^2\left(\frac{k^2_{1 \bot}}{\mu^2}\right)\right]\right\}. 
\end{equation}

Form factor $f_n$ related to emission of $n$ gluons is given by the similar formula: 

\begin{equation}\label{fn}
f_n \left(\frac{s}{\mu^2}, \frac{k^2_{1 \bot}}{\mu^2}\right) = \exp \left\{- \frac{\alpha_s}{2 \pi}\left[C_F \ln^2\left(\frac{s}{\mu^2}\right) + 
\frac{N}{2}\sum_{i = 1}^n \ln^2\left(\frac{k^2_{i \bot}}{\mu^2}\right)\right]\right\}. 
\end{equation}

The quark masses in Eqs.~(\ref{f1},\ref{fn}) are neglected. In order to account for them, one needs 
substitute $F_0$ in Eq.~(\ref{matchf1f0}) by $F_m$ of  Eq.(\ref{f0wm}). Expressions for $f_n$ with off-shell quarks 
can be obtained in the way altogether similar to Sect.~IV.

Expressions for $f_n$ of Eq.~(\ref{fn}) were used in Refs.~\cite{df} for analyses of parton jets and soft hadron spectra. 
To complete discussing Sudakov form factors, we note that $f(q^2)$ in the skewed regime was considered in Ref.~\cite{kimpiv}.

\section{Form factor $g$ in DLA}

 The electromagnetic vertex $\Gamma_{\mu}$ of Eq.~(\ref{gammadef})  involves form factors $f(q^2)$ and $g(q^2)$. 
In the previous Sects. we considered DL asymptotics of the form factor $f(q^2)$ in QED, QCD and theory of EW interactions.
Now we move on to DL asymptotics of $g(q^2)$. Below we briefly reproduce results of Ref.~\cite{etg}. 
We start by considering $g(q^2)$ in QED and then proceed to QCD.
%
%
For the sake of simplicity we assume that both the incoming and outgoing electrons are on-shell. Form factors $f$, $g$ are 
quite different in low orders. Indeed, the Born approximation yields that $f^{Born} = 1$ while $g^{Born} = 0$. Then, the 
first-loop contributions to $f$ and $g$ are also widely different. The leading contribution to $f(s)$ is double-logarithmic:

\begin{equation}\label{f1dl}
f^{(1)} (s) = - \frac{\alpha}{4 \pi} \ln^2 \left(\frac{s}{\mu^2}\right),
\end{equation}
with $\mu$ being IR cut-off, 
while the leading contribution to $g(s)$ is single-logarithmic and power-suppressed\cite{shw}:

\begin{equation}\label{g1dl}
g^{(1)} \left(\frac{s}{m^2}\right) = - \frac{\alpha}{\pi} \;\frac{m^2}{s} \ln \left(\frac{s}{m^2}\right),
\end{equation}
with $m$ being the electron mass. There is a deep qualitative difference between parameters $\mu$ and $m$ in 
logarithms in Eqs.~(\ref{f1dl},\ref{g1dl}). 
Indeed, $\mu$ in Eq.~(\ref{f1dl}) is the IR cut-off. Its value is arbitrary and exploiting this fact makes possible to obtain 
$f^{(1)} (s)$ and other DL 
contributions to $f (s)$ by tracing their evolution with respect to $\mu$ and using $f^{Born}$ as the starting point.
 On the contrary, $g^{(1)} (s)$ is IR stable,  
$m$ is fixed, so $g^{(1)} (s)$ cannot be obtained by any evolution. Nevertheless, higher loop contributions to $g$ yield IR divergencies, so 
introducing the IR cut-off $\mu$ becomes mandatory while $g^{(1)} (s)$ acts as the starting, "Born" term. In other words, 
the difference between IREEs for $g (s)$ and $f (s)$ is in the starting points only. IREE for $g (s)$ is depicted in Fig.~8. 
It can be constructed and solved in exactly the same way as Eq.~(\ref{eqf0k}). 
Therefore, IREE for $g (s)$ is 

\begin{eqnarray}\label{eqg}
g &=& g^{(1)}\left(\frac{s}{m^2}\right) - \frac{\alpha_s C_F}{2 \pi} \int_{\mu^2}^{s} \frac{d k^2_{\perp}}{k^2_{\perp}} 
\ln \left(\frac{s}{k^2_{\perp}}\right) \;g\left(\frac{s}{k^2_{\perp}}\right).
\end{eqnarray}

Differentiating it with respect to $\mu$ leads to the differential equation
\begin{equation}\label{eqgdif}
\frac{d g(\rho)}{d \rho} = - \frac{\alpha_s C_F}{2 \pi} \rho g(\rho),
\end{equation}
with the obvious solution 

\begin{equation}\label{gmu}
 g(\rho) = g^{Born}(s/m^2) \exp\left[- \frac{\alpha_s C_F}{4 \pi} \rho^2\right]
\end{equation}

It is interesting that if we put $\mu = m$ in Eq.~(\ref{gmu}), it can be written in a more elegant way:

\begin{equation}\label{gm}
 g(\xi) = 
- 2\frac{d f (\xi)}{d \xi},
\end{equation}
with $\xi = s/m^2$, and therefore the asymptotics of the vertex $\Gamma_{\mu}$ in DLA can be written as follows:

\begin{equation}\label{gammaxi}
\Gamma_{\mu} (\xi) = \left[\gamma_{\mu} + \frac{\sigma_{\mu \nu} q_{\nu}}{m} \frac{d}{d \xi}\right] f(\xi).
\end{equation}

Similarly to the situation with $f(s)$, the expression of $g(s)$ in QCD can be obtained from Eq.~(\ref{gmu}) with the  
replacement of $\alpha$ by $\alpha_s C_F$.

\section{Applications of DLA to processes in Regge kinematics}

Exponential (Sudakov) suppression  of the form factors considered in the previous Sects. is also true for the 
scattering amplitudes of $2 \to 2$ reactions of the type 
$a(p), b(q) \to c(p^{\prime}), d(q^{\prime})$ 
at high energies in the hard kinematics, where 

\begin{equation}\label{hardkin}
s = (p + q)^2 \sim -u = - (p - q^{\prime}) \sim - t = - (p^{\prime} - p)^2
\end{equation}
and DL contributions in this kinematics  arrive from soft bosons (photons and gluons). 
However, it does not stand for the scattering amplitudes in Regge kinematics, 
where DL contributions of the Sudakov type arriving from soft bosons (photons and gluons) 
are complemented with DLs from $t$-channel states made of soft fermion (leptons and quarks) pairs as 
proved in Refs.~\cite{ggfl1, ggfl2}. 
There are two types of Regge kinematics for $2 \to 2$ reactions: \\
forward kinematics, where 

\begin{equation}\label{fkin}
s \sim -u \gg - t
\end{equation}
and backward kinematics, where 

\begin{equation}\label{bkin}
s \sim -t \gg - u.
\end{equation}

Amplitudes of the forward and backward $e^+ e^- \to \mu^+ \mu^-$ -annihilation were calculated in DLA in Refs.~\cite{ggfl1, ggfl2} 
and generalisation of these results to QCD was done in Ref.~\cite{kl}, where the first IREEs were constructed. Since then DLA 
was used to calculate many objects at high energies. In the present paper, we restrict ourself by discussing two 
cases only: elastic photon-photon scattering 
$\gamma^{\star}(p), \gamma^{\star}(q) \to \gamma^{\star}(p^{\prime}), \gamma^{\star}(q^{\prime})$ and DIS. 

Amplitude of $\gamma^{\star}(p), \gamma^{\star}(q) \to \gamma^{\star}(p^{\prime}), \gamma^{\star}(q^{\prime})$-
forward scattering  via a single quark loop was calculated with DL accuracy at $t = 0$ in Refs.~\cite{bl1,bl2}, where different 
methods, including IREE, were considered and all of them yielded the same result, which confirmed validity and efficiency 
of the IREE approach.  

Application of DLA to Polarized DIS:  
Non-singlet component $F_1^{NS}$ to the DIS structure function $F_1$ was calculated in DLA in Ref.~\cite{emr}. There 
was obtained the expression for $F_1^{NS}$ and calculated its small-$x$ asymptotics which proved to be of the Regge 
 type, i.e. $F_1^{NS} \sim x^{- \omega_0^{NS}}$, with 
the intercept 

\begin{equation}\label{intfns}
\omega_0^{NS} = \sqrt{2 \alpha_s C_F/\pi}.
\end{equation}

This calculation demonstrated that $F_1^{NS}$ grows much faster at small $x$ than it is prescribed by the DGLAP equations. 
Eq.~(\ref{intfns}) was confirmed in Ref.~\cite{kpsfns}, where the method alternative to IREE was used. \\

Then  the non-singlet $g_1^{NS}$ and singlet $g_1^{S}$ components of the spin structure function $g_1$ 
 were 
calculated with DL accuracy in Refs.~\cite{berns,bers}. 
Asymptotics of these structure functions also proved to be of the Regge type. The non-singlet intercept $\Delta_{NS}$ was calculated 
analytically, though approximately: 

\begin{equation}\label{deltans}
\Delta^{NS} = \omega_0^{NS} \sqrt{\frac{1 + \sqrt{1 + 4/(N C_F)}}{2}} \approx \omega_0^{NS} \left[1 + 1/2N^2\right].
\end{equation}

In contrast, the singlet intercept was calculated by numerical means: 

\begin{equation}\label{deltas}
\Delta_s = 3.45 \sqrt{\frac{\alpha_s N}{2 \pi}}.
\end{equation}

When quark contributions are dropped, the intercept $\widetilde{\Delta}_s$ is larger: 

\begin{equation}\label{deltatildes}
\widetilde{\Delta}_s = 3.66 \sqrt{\frac{\alpha_s N}{2 \pi}}.
\end{equation}

This result was confirmed in Ref.~\cite{kpsgs}, where the KPSCTT approach was used instead of IREE. \\

A quite important application of DLA  to study parton jets was presented in Refs.~\cite{doktr1,catdok1,catdok2}. 
A part of single-logarithmic contributions 
to parton jets was added to DL terms in Ref.~\cite{doktr2}. 

Although this recap of application of DLA to high-energy processes is not complete, it demonstrates how many 
interesting results were obtained in the DLA framework.

\section{Conclusion}

We started the present paper by reminding how DL coming from virtual soft bosons (photons) were discovered in the pioneer paper\cite{sud}. 
However, such bosons were not the only source of DL. 
It was shown in Refs.~\cite{ggfl1,ggfl2,ggfl4}  that DL come also from 
pairs of virtual soft fermions and there are no other sources of DL. Amplitudes of forward and backward 
$e^+e^- \to \mu^+ \mu^-$ -annihilation were calculated in Refs.~~\cite{ggfl1,ggfl2,ggfl4} by the quite cumbersome methods which 
scarcely could be applied to more complicated processes. Fortunately, a much more efficient approach was suggested on the basis 
of factorization of photons with minimal $k_{\bot}$  proved in Ref.~\cite{g}. 
The factorization was applied in Ref.~\cite{kl} to elastic quark-antiquark annihilation, which eventually led to creating the new method 
of calculations with DL accuracy: composing Infra-Red Evolution Equations. This method proved to be both a simple and efficient instrument. It has been applied to various reactions 
high-energy processes as elastic as inelastic. We have given several examples of them in the present paper, 
focusing on early publications, but actually this list is far 
from a complete set of references. \\

\section{Acknowledgement}

I am grateful to M.G.~Ryskin and V.A.~Khoze for useful communications. \\

\textbf{DECLARATION of COMPETING INTEREST}\\

The author declares that he have no known competing financial interests or personal relationships that could have appeared to 
influence the work reported in this paper\\

\textbf{DATA AVAILABILITY}\\

No data was used for the research described in the article\\


%
%


\begin{thebibliography}{99}

\bibitem{sud} V.V.~Sudakov. Sov. Phys. JETP 3(1956)65.

\bibitem{abr} A.A.~Abrikosov. Zh. Exp. Theor Phys. 30, 386, 544, 1956. 

\bibitem{mil} G.A.~Milekhin, E.S.~Fradkin. Zh. Exp. Theor Phys. 45, 1926, 1963. 

\bibitem{ggfl1} V.G.~Gorshkov, V.N.~Gribov, G.V.~Frolov, L.N.~Lipatov. Sov.J.Nucl.Phys. 6 (1968) 95.

\bibitem{ggfl2} V.G.~Gorshkov, V.N.~Gribov, G.V.~Frolov, L.N.~Lipatov. Sov.J.Nucl.Phys. 6 (1968) 262. 

\bibitem{ggfl3} V.G.~Gorshkov, V.N.~Gribov, G.V.~Frolov, L.N.~Lipatov. Annals Phys. 43 (1967) 201. 

\bibitem{ggfl4} V.G.~Gorshkov, V.N.~Gribov, G.V.~Frolov, L.N.~Lipatov. Phys.Lett. 22 (1966) 671. 

\bibitem{khoze} Ya.I.~Azimov, A.I.~Vainshtein, L.N.~Lipatov, V.A.~Khoze. JETP Lett. 21 (1975) 172.

\bibitem{g} V.N.~Gribov. Sov.J.Nucl.Phys. 5 (1967) 280. 

\bibitem{kl} R.~Kirschner, L.N.~Lipatov.     Sov.Phys.JETP 56 (1982) 266;  Nucl.Phys.B 213 (1983) 122. 

\bibitem{dglap} G.~Altarelli and G.~Parisi, Nucl.~Phys.B126 (1977) 297;
V.N.~Gribov and L.N.~Lipatov, Sov.~J.~Nucl.~Phys. 15 (1972) 438;
L.N.Lipatov, Sov.~J.~Nucl.~Phys. 20 (1972) 95; Yu.L.~Dokshitzer,
Sov.~Phys.~JETP 46 (1977) 641.

\bibitem{emr} B.I.~Ermolaev, S.I.~Manaenkov, M.G.~Ryskin. Z.Phys.C 69 (1996) 259.

\bibitem{berns} J.~Bartels, B.I.~Ermolaev, M.G.~Ryskin. Z.Phys.C 70 (1996) 273. 

\bibitem{bers} J.~Bartels, B.I.~Ermolaev, M.G.~Ryskin.     Z.Phys.C 72 (1996) 627.

\bibitem{el} B.I.~Ermolaev, L.N.~Lipatov. Int.J.Mod.Phys.A 4 (1989) 3147.

\bibitem{egtalpha} B.I.~Ermolaev, M.~Greco, S.I.~Troyan.     Riv.Nuovo Cim. 33 (2010) 2, 57. 

\bibitem{quinn} G.~Carazzone, E.C.~Poggio, H.R.~Quinn.     Phys.Lett.B 57 (1975) 161. 

\bibitem{smilga} A.V.~Smilga. Phys.Lett.B 83 (1979) 357. 

\bibitem{tikt} J.M.~Cornwall, G.~Tiktopoulos.     Phys.Rev.D 13 (1976) 3370. 

\bibitem{flmm} V.S.~Fadin, L.N.~Lipatov, A.D.~Martin, M.~Melles. Phys. Rev. D61 (2000) 094002. 

\bibitem{low1} F.F.~Low. Phys.Rev. 96 (1954) 1428. 

\bibitem{low2} M.~Gell-Mann, M.L.~Goldberger. Phys.Rev. 96 (1954) 1433. 

\bibitem{l} L.N.~Lipatov.  Nucl.Phys.B 307 (1988) 705; Nucl.Phys.B 307 (1988) 705. 

\bibitem{ce} M.~Chaihian, B.~Ermolaev.     Nucl.Phys.B 451 (1995) 194.
 
\bibitem{kf} E.A.~Kuraev, V.S.~Fadin. Yad.Fiz. 27 (1978) 1107. 

\bibitem{efl} B.I.~Ermolaev, V.S.~Fadin, L.N.~Lipatov.     Yad.Fiz. 45 (1987) 817. 

\bibitem{kurcher}  E.~Bartos, E.A.~Kuraev, I.O.~Cherednikov.   Phys.Lett.B 593 (2004) 115.

\bibitem{df} Y.L.~Dokshitzer, V.S.~Fadin, V.A.~Khoze.     
Phys.Lett.B 115 (1982) 242;     Z.Phys.C 15 (1982) 325; Z.Phys.C 18 (1983) 37. 

\bibitem{kimpiv} V. T.~Kim, V. A.~Matveev, G. B.~Pivovarov.     Phys.Rev.D 99 (2019) 2, 025016.	

\bibitem{shw}  J.Schwinger. Phys. Rev. 73 (1948) 416.

\bibitem{etg} B.I.~Ermolav, S.I.~Troyan. Nucl.Phys.B 590 (2000) 521.

\bibitem{bl1} J. Bartels, M. Lublinsky, JHEP 0309 (2003) 076. 

\bibitem{bl2} J.~Bartels, M.~Lublinsky. Mod. Phys. Lett. A 19 (2004) 19691982.



\bibitem{kpsfns} Yuri V. Kovchegov, Daniel Pitonyak, Matthew D. Sivert. Phys.Rev.D 95, 014033 (2017).

 

\bibitem{kpsgs} F. Cougoulic, Y. V. Kovchegov, A. Tarasov and Y. Tawabutr. JHEP 07, 095 (2022).

\bibitem{doktr1} Yu.L.~Dokshitzer, V.A.~Khoze, S.I.~Troian. J.Phys.G 17 (1991) 1481. 

\bibitem{catdok1} S.~Catani, Yu.L.~Dokshitzer, B.R.~Webber.     Phys.Lett.B 322 (1994) 263.

\bibitem{catdok2} S.~Catani, Yu.L.~Dokshitzer, B.R.~Webber. Nucl.Phys.B 383 (1992) 419.

\bibitem{doktr2} Yu.L.~Dokshitzer, V.A.~Khoze, S.I.~Troian.     Z.Phys.C 55 (1992) 107. 

\end{thebibliography}
\end{document}